\documentclass[aps,prl,amsmath,amssymb,superscriptaddress,showpacs,twocolumn]{revtex4-1}

\usepackage{graphicx} \usepackage{color} \usepackage{verbatim}

\def\vs{{\it vs.}\,}
\def\rhoM{\rho_{\mathrm{MIR}}}
\def\TM{T_{\mathrm{MIR}}}
\def\TFL{T_{\mathrm{FL}}}
\def\Tknee{T_{*}}
\def\Imag{\mathrm{Im}}
\def\Real{\mathrm{Re}}
\def\isigma{\mathrm{Im}\Sigma}
\def\rsigma{\mathrm{Re}\Sigma}
\def\ek{\varepsilon_{\mathbf{k}}}
\def\eF{\varepsilon_F}
\def\Ak{A_{\mathbf{k}}(\omega)}
\def\omm{\omega_{-}}
\def\omp{\omega_{+}}
\def\chiloc{\chi_{\mathrm{loc}}}

\def\K{\mathrm{K}}

\def\frate{Fig.~\ref{fig:rate}\,}
\def\fopt{Fig.~\ref{fig:optics}\,}
\def\fth{Fig.~\ref{fig:thermo}\,}

\def\LiVO{LiV$_2$O$_4$\,}

\def\SROtwo{Sr$_{2}$Ru$_{}$O$_{4}$\,}

\begin{document}

\title{How bad metals turn good: spectroscopic signatures of resilient quasiparticles}

\author{Xiaoyu Deng}
\affiliation{Centre de Physique Th\'eorique,
  Ecole Polytechnique, CNRS, 91128 Palaiseau Cedex, France}
\affiliation{Japan Science and Technology Agency, CREST, Kawaguchi 332-0012, Japan} 
\affiliation{Department of Physics, Rutgers University, Piscataway, NJ 08854, USA}
\author{Jernej Mravlje}
\affiliation{Coll\`ege de France, 11 place Marcelin Berthelot, 75005 Paris, France}
\affiliation{Centre de Physique Th\'eorique, Ecole Polytechnique, CNRS, 91128 Palaiseau Cedex, France}
\affiliation{Jo\v{z}ef Stefan Institute, Jamova~39, Ljubljana, Slovenia}
\author{Rok \v{Z}itko}
  \affiliation{Jo\v{z}ef Stefan Institute, Jamova~39, Ljubljana, Slovenia}
\author{Michel Ferrero}
\affiliation{Centre de Physique Th\'eorique, Ecole Polytechnique, CNRS, 91128 Palaiseau Cedex, France}
\author{Gabriel Kotliar}
\affiliation{Department of Physics, Rutgers University, Piscataway, NJ 08854, USA}
\author{Antoine Georges}
 \affiliation{Coll\`ege de France, 11 place Marcelin Berthelot, 75005 Paris, France}
 \affiliation{Centre de Physique Th\'eorique, Ecole Polytechnique, CNRS, 91128 Palaiseau
    Cedex, France}
 \affiliation{DPMC, Universit\'e de Gen\`eve, 24 quai Ernest Ansermet, CH-1211 Gen\`eve, Suisse}
 \affiliation{Japan Science and Technology Agency, CREST, Kawaguchi
    332-0012, Japan}

\date{\today}

\begin{abstract}
We investigate transport in strongly-correlated metals.  
Within dynamical mean-field theory, we calculate the resistivity, thermopower, 
optical conductivity and thermodynamic properties of a hole-doped Mott insulator. 
Two well-separated temperature scales are identified: $\TFL$ below which Landau Fermi liquid behavior applies, 
and $\TM$ above which the resistivity exceeds the Mott-Ioffe-Regel value and `bad-metal' behavior is found. 
%
We show that quasiparticle excitations remain well-defined above $\TFL$ and dominate transport throughout the intermediate regime $\TFL \lesssim T \lesssim \TM$. 
%
The lifetime of these `resilient quasiparticles' is longer for electron-like excitations, and this pronounced particle-hole asymmetry has
important consequences for the thermopower.
The crossover into the 
bad-metal regime corresponds to the disappearance of these excitations, and has clear signatures in  optical spectroscopy. 
\end{abstract}

\maketitle

The transport properties of metals with strong electron correlations
are %
unconventional and 
poorly understood theoretically. 
Two %
facts regarding the temperature
dependence of the resistivity %
are frequently observed.
(i) Fermi-liquid (FL) behavior $\rho\propto T^2$
only holds below a
temperature $\TFL$ which is %
low compared to bare electronic energy scales. 
(ii) 
At high temperatures the resistivity is large and reaches %
values exceeding the Mott-Ioffe-Regel (MIR) value.
This `bad-metallic' behavior~\cite{emery_kivelson_prl_1995} signals
the breakdown of a quasiparticle (QP) description of transport, since the
associated mean-free path $l$ would be smaller than the lattice spacing.
This is observed in many materials: \SROtwo has
$\TFL\simeq 20\K$, while the MIR value is reached at $\TM\simeq 800\K$
(using $k_F l\sim 1$ as the MIR criterion); in \LiVO, $\TFL$ is a few
degrees Kelvin, while $\TM$ is several hundreds,
etc. (see~\cite{hussey_phil_mag_2004,gunnarsson_rmp_2003} for reviews).
In two-dimensional organic materials, $\TFL$ and $\TM$ are closer but still 
distinct scales~\cite{merino_mckenzie_prb_2000,limelette_prl_2003,merino_prl_2008}. 

These observations raise  the following questions. 
Why is $\TFL$ much lower than $\TM$ and what determines its value?
Up to which temperature do QPs exist and what are the signatures of their disappearance? 
And, most importantly: how should one think of transport not only in
the bad-metal, but also in the intermediate regime $\TFL \lesssim T
\lesssim \TM$ where the resistivity does not follow Landau's
$T^2$ behavior, but is still smaller than the MIR value?
These questions also apply to cuprate superconductors,
where the observation of quantum oscillations~\cite{Hussey_Nature_2003,doiron_proust_nature_2007} and 
$T^2$ behavior in transport~\cite{hussey_jphys_2008,barisic_12} and optics~\cite{mirzaei_vdMarel_2012} 
have rejuvenated the relevance of FL states with a low $\TFL $ (possibly 
with angular dependence along the Fermi surface).

In this article, we answer these questions in a particularly simple setting: 
a hole-doped Mott insulator described with dynamical mean-field theory (DMFT).
Our most striking finding is that {\it 
well-defined QP excitations survive well above the range of validity of FL theory}.
%
%
%
%
%
%
%
%
%
For over a decade in temperature above $\TFL$ the transport can be accurately described in terms of these `resilient quasiparticles' (RQPs).
The high-temperature MIR crossover into the bad-metal regime is
associated with their gradual extinction, which has  a
clear signature both in the single-particle spectral function and in optical spectroscopy.
In a hole-doped Mott insulator, the RQPs come with a 
strong particle-hole asymmetry: electron-like excitations are 
longer-lived than hole-like ones.
This has direct consequences for the thermopower.%

Previous DMFT work has investigated transport~\cite{pruschke93, jarrell_prb_1994,kajueter_prb_1996,
palsson_prl_1998,palsson_thesis_2001} 
and optical conductivity~\cite{jarrell_prb_1995,merino_prl_2008}, but not the 
precise
temperature-dependence 
of the self-energy and  of the momentum-resolved spectral function which reveals this intermediate RQP regime. 
We note that in the half-filled case relevant to organic
compounds~\cite{merino_mckenzie_prb_2000,limelette_prl_2003}, the
high-temperature state is insulating-like, and hence the temperature
window where this regime can be seen is narrower. 
We solved the DMFT equations~\cite{georges_rmp_1996} for the hole-doped Hubbard model 
using highly accurate numerical-renormalization group (NRG)~\cite{bulla_rmp_2008,NRGLjubljana}
and continuous-time quantum Monte Carlo~\cite{gull_rmp_2011,TRIQS} techniques.
In DMFT the real part of the optical conductivity reads
\begin{equation}
\sigma(\omega)= \frac{2\pi e^2}{\hbar} \int d\omega'
F_{\omega, \omega'} 
 \int d\epsilon \Phi(\epsilon)  A_{\mathbf k} (\omega') A_{\mathbf k} (\omega'+\omega) 
 \label{eq:opt_con}
\end{equation}
where $F_{\omega,\omega'}=[f(\omega')-f(\omega+\omega')]/\omega$ with 
$f(\omega)$ the Fermi function.
$\Ak=-(1/\pi) \mathrm{Im} [\omega + \mu - \ek -\Sigma(\omega)]^{-1}$ is the 
single-particle spectral function, with $\ek$ the energy of the state in the band,
$\Sigma(\omega)$ the retarded self energy and $\mu$ the chemical potential. 
$\Phi(\varepsilon) = (1/V)\sum_{\mathbf k} (\partial\ek/\partial{\mathbf k}_x)^2
\delta(\varepsilon-\ek)$ contains the information about velocities. 
We used a semicircular density-of-states (DOS) with a half-width $D$, and the corresponding sum-rule preserving 
expression $\Phi(\epsilon)=\Phi(0)\left[1-(\epsilon/D)^2\right]^{3/2}$. 
In the following, resistivity will be expressed in units of the MIR value defined as 
$1/\rho_\mathrm{MIR}\equiv e^2 \Phi(0)/\hbar D$. 
This choice is consistent with the criterion $k_F l=1$ for a parabolic band in two dimensions,  
for which the conductivity $\sigma=(k_F l)\,e^2 \Phi(\eF)/\hbar \eF=(k_Fl)e^2/h$. 
\begin{figure}
\includegraphics[width=1\columnwidth]{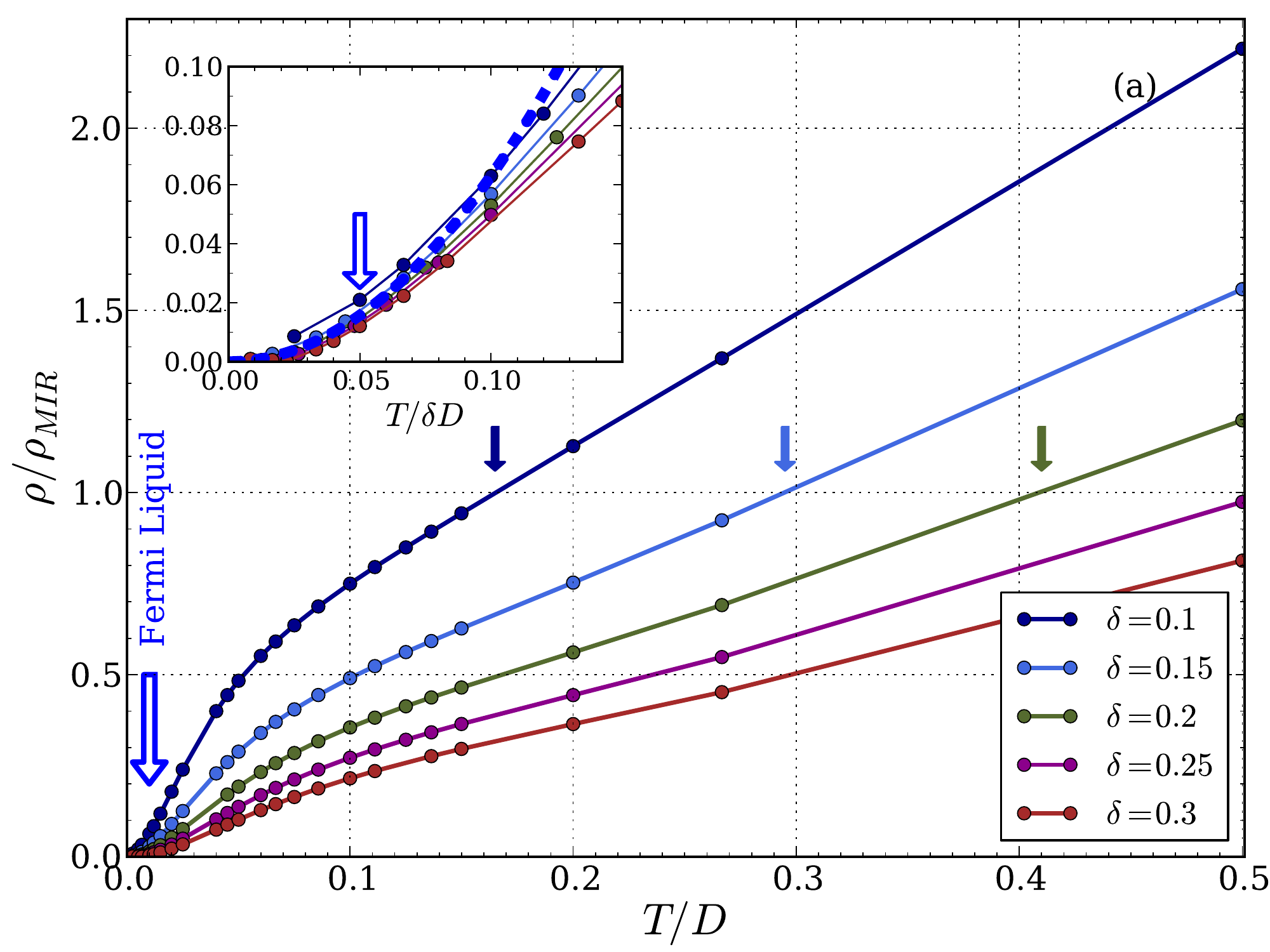}
\includegraphics[width=1\columnwidth]{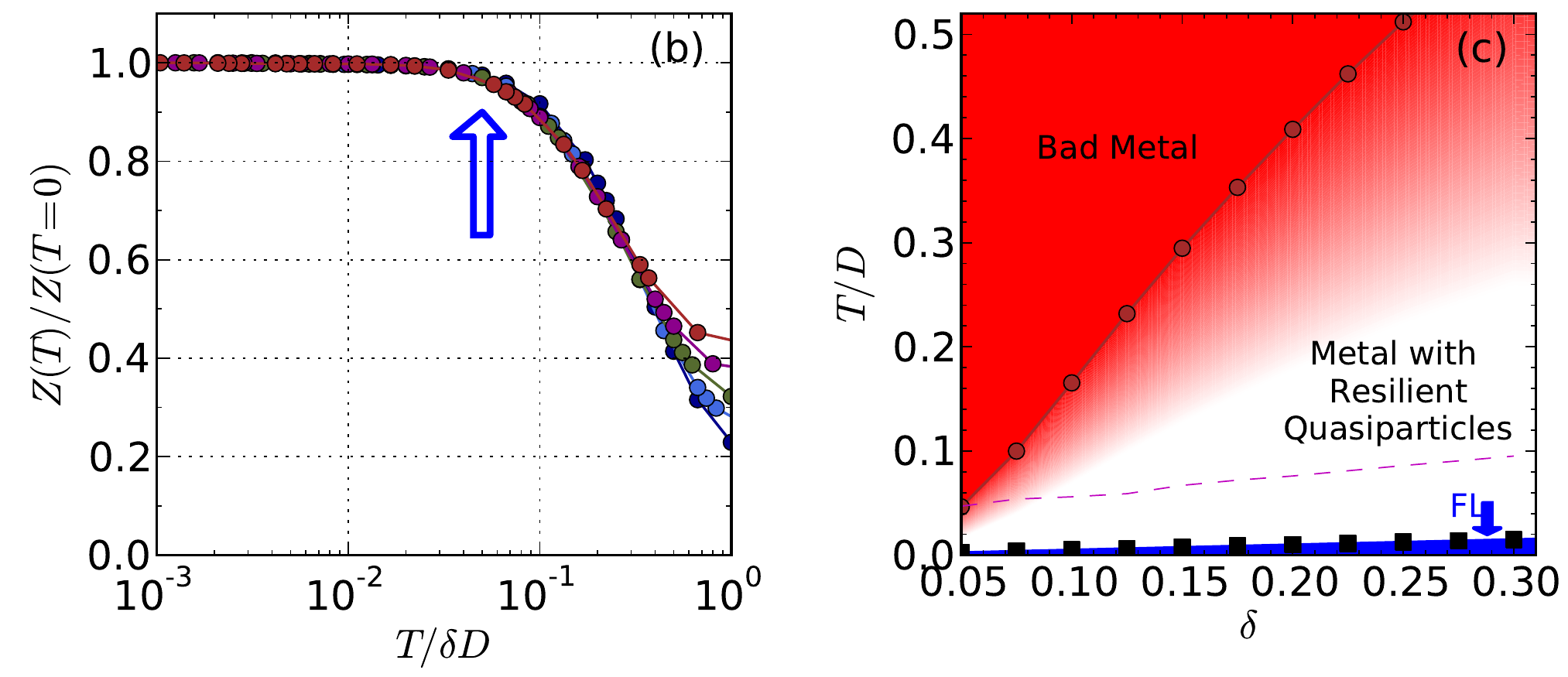}
  \caption{(a) Temperature-dependence of the resistivity for several doping levels $\delta$ 
  ($U/D=4$, as in all figures). 
  The MIR value as defined in the text is reached at a temperature $\TM$ indicated by plain arrows.  
   Inset: resistivity at low temperatures \vs $T/\delta D$ revealing the %
$T^2$ behavior (dashed line) 
   below $\TFL$ (empty arrow). 
 (b)  Determination of $\TFL$ by a scaling plot of $Z(T)/Z(T\rightarrow 0)$ \vs $T/\delta D$. 
       Here, $Z(T)^{-1}\equiv 1-\partial\Real\Sigma(\omega,T)/\partial\omega$.  
 (c)  The different regimes: FL (blue) for $T<\TFL$, bad metal (red) and intermediate RQP regime.
  The crossover into the bad metal is gradual: the onset of red shading corresponds to the 
  optical spectroscopy signatures discussed in the text, while the red points indicate where 
  $\rhoM$ is reached.
  The thin dashed line indicates the `knee' in $\rho(T)$.  
\label{fig:rho} }
\end{figure}

Fig.~\ref{fig:rho}c summarizes our main result: as a function of temperature,
three distinct regions appear. At low $T < \TFL$, FL behavior is found. At high
temperature, the system is a bad metal with no quasiparticles (this is
indicated by the shaded area on Fig.~\ref{fig:rho}c, more details below).
Between these two limits, there is an extended region with well-defined
quasiparticles, but that do not obey FL behavior. The nature of this metal with
`resilient quasiparticles' (RQPs) is the central focus of our work.

Let us first discuss transport (resistivity \vs $T$, Fig.~\ref{fig:rho}a)
in the light of these regimes. At low temperature the
resistivity has a FL $T^2$ behavior. This extends up to a temperature $\TFL$
(see inset in Fig.~\ref{fig:rho}a) which is proportional to the doping level,
$\TFL\,\simeq 0.05 \delta D$. Note that $\TFL$ can be determined by other
complementary criteria, like the scaling of $\Imag\Sigma(\omega,T)/T^2$ vs.
$\omega/T$ (see on-line supplementary~\cite{supp}) or the scaling of
$Z(T=0)/Z(T)$ vs. $T/\delta D$ (Fig.~\ref{fig:rho}b).
Above $\TFL$, the resistivity increases approximately linearly (with a negative
intercept). A knee-like feature is observed at a temperature $\Tknee$, above
which the high-$T$ regime gradually sets in. $\rho(T)$ then has a linear
temperature dependence (with a positive intercept), as can be shown from a
high-$T$ expansion~\cite{palsson_prl_1998,palsson_thesis_2001}, and smoothly
crosses $\rhoM$ (see arrows in Fig.~\ref{fig:rho}a) at a temperature $\TM \sim
2\delta D$.

\begin{figure}
\includegraphics[width=1\columnwidth]{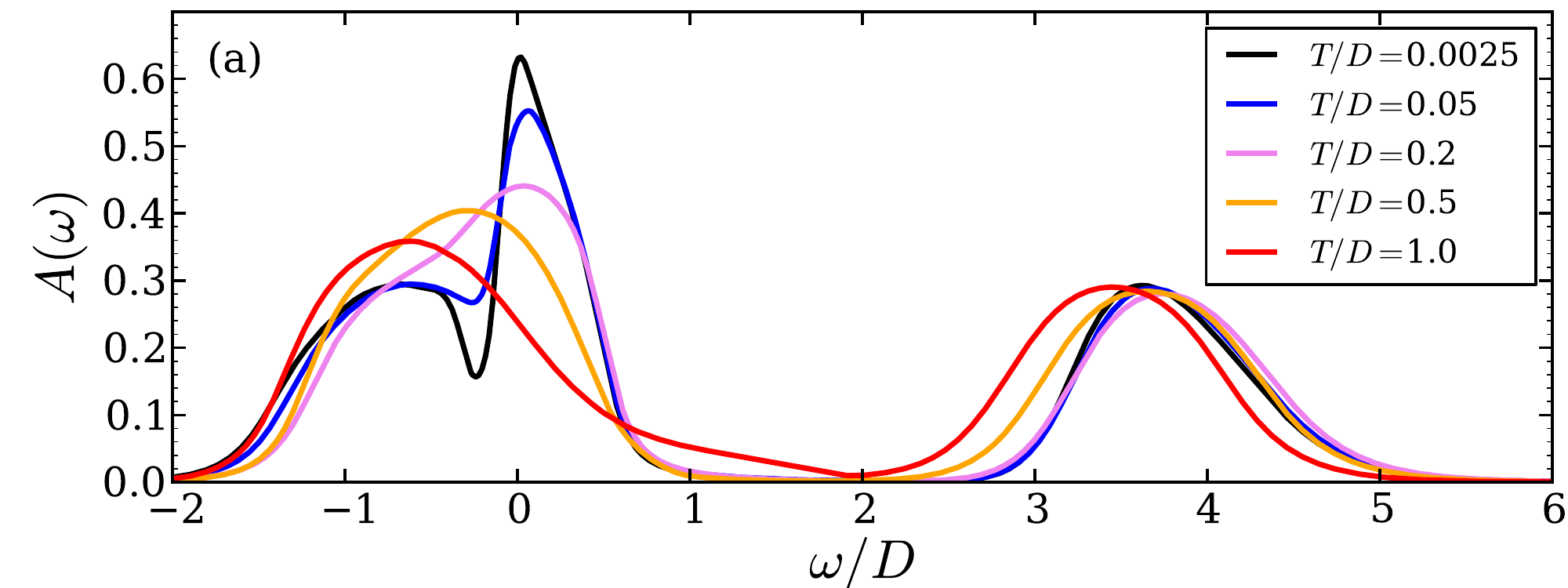}
\includegraphics[width=1\columnwidth]{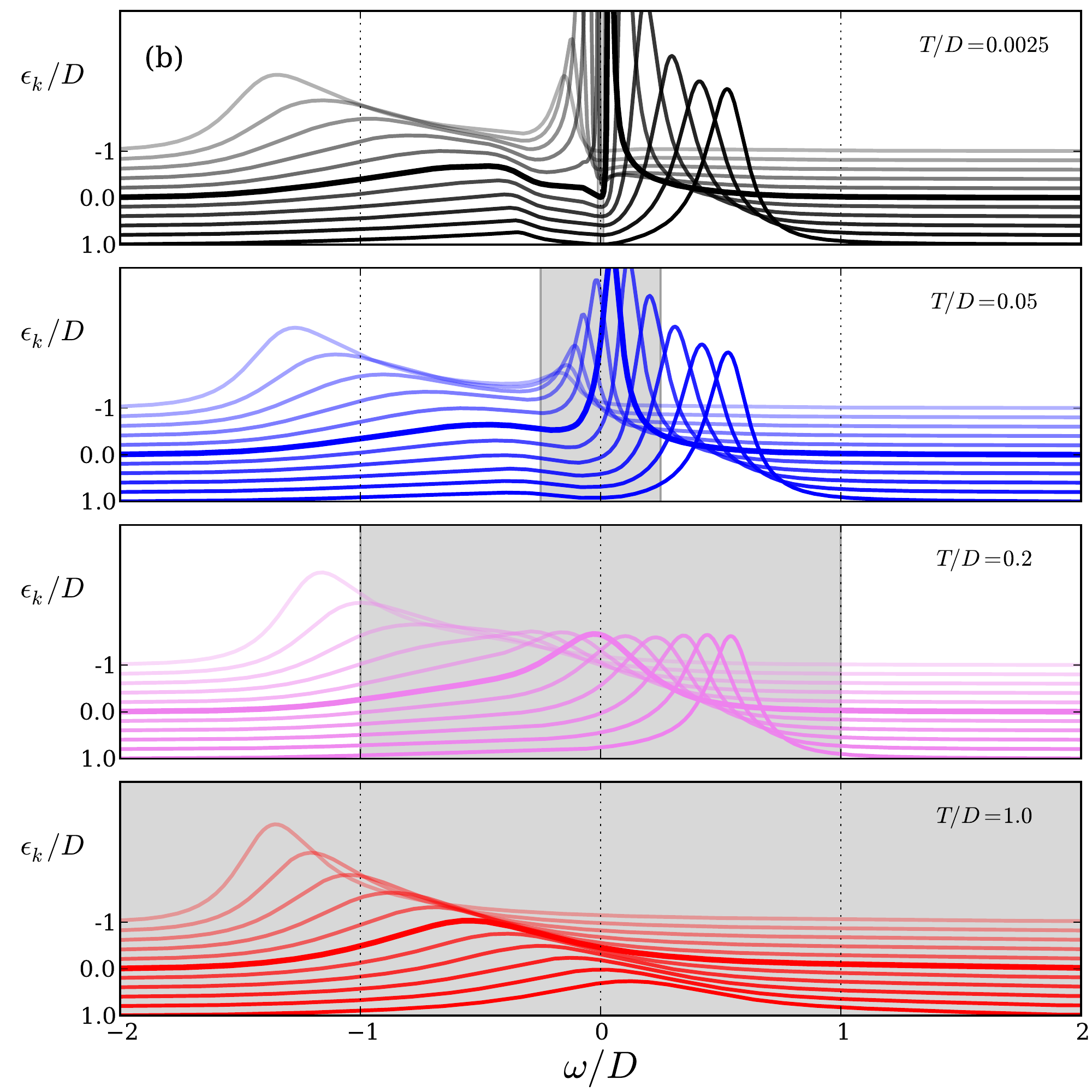}
\caption{(a) Temperature evolution of the total DOS for $\delta=0.2$. 
(b) Momentum- resolved spectral functions. The shaded area [-5$k_BT$,5$k_BT$] indicates the states with a 
significant contribution to transport.
\label{fig:spectra}}
\end{figure}

The data therefore show that there is a
wide temperature range in which transport does not follow the $T^2$ FL
behavior, although the resistivity is still substantially smaller than
$\rhoM$. This raises the following question: what
are the charge carriers in this intermediate metallic regime? 
To this aim, we depict in
Fig.~\ref{fig:spectra}
the momentum-resolved spectral function $\Ak$ at selected temperatures as energy distribution curves
($\Ak$ \vs $\omega$ for different $\ek$'s as well as the momentum-integrated DOS).
These results reveal a remarkable fact:
well-defined QP excitations exist throughout this intermediate regime, way above the FL scale.
Our definition of the term 'quasiparticle' is a pragmatic one: we mean
that $\Ak$ displays  a well-resolved peak in the vicinity of the Fermi level, in addition to a lower Hubbard band (LHB) and an upper Hubbard band (UHB).

For $T<\TFL$ ($T/D=0.0025$ in Fig.~\ref{fig:spectra}) %
sharp peaks are seen close to the Fermi energy ($\omega=0$), characteristic of long-lived Landau QPs.
For $T>\TFL$ ($T/D=0.05$ curves), the peaks broaden and the RQPs are visible mostly on the 
unoccupied side of the FS ($\omega>0$).
As the temperature increases (see e.g. $T/D=0.2$), the QP can barely be
resolved (see~\cite{supp}, for a color map representation) and eventually
disappear, with just the Hubbard satellites remaining in the spectra (e.g. the
two-peak structure that is visible in the total DOS of Fig.~\ref{fig:spectra}a
at $T/D=0.2$ is not present for $T/D=0.5$ anymore). This crossover into the
bad-metallic regime is a gradual one and there is not a precise temperature
where the QP suddenly die out (below we discuss how the optical conductivity
provides a criterion for the onset of the bad-metallic behavior). Our data
nevertheless clearly show that they still exit well above $\Tknee$ and that they
have completely disappeared at $\TM$.

$\TFL$ and $\TM$ appear as overall scales between which RQPs exist. Both these
temperatures are proportional to the doping level $\delta$ but with very
different prefactors $\TFL / \TM \simeq 0.025$. Correspondingly, the 
resistivity at $\TFL$ is much smaller than $\rhoM$, $\rho(\TFL)/\rhoM \simeq
0.016$ (a low-$T$ expansion in the FL region yields $\rho(T)/\rhoM \sim 6.3
(T/\delta D)^2 +\cdots$). Let us emphasize that the Brinkman-Rice scale $\delta
D$, which is a measure of the kinetic energy of QPs and hence of the quantum
degeneracy scale, is associated with $\TM$, not with $\TFL$.

%
%
%
%
\begin{figure}
\includegraphics[width=1\columnwidth]{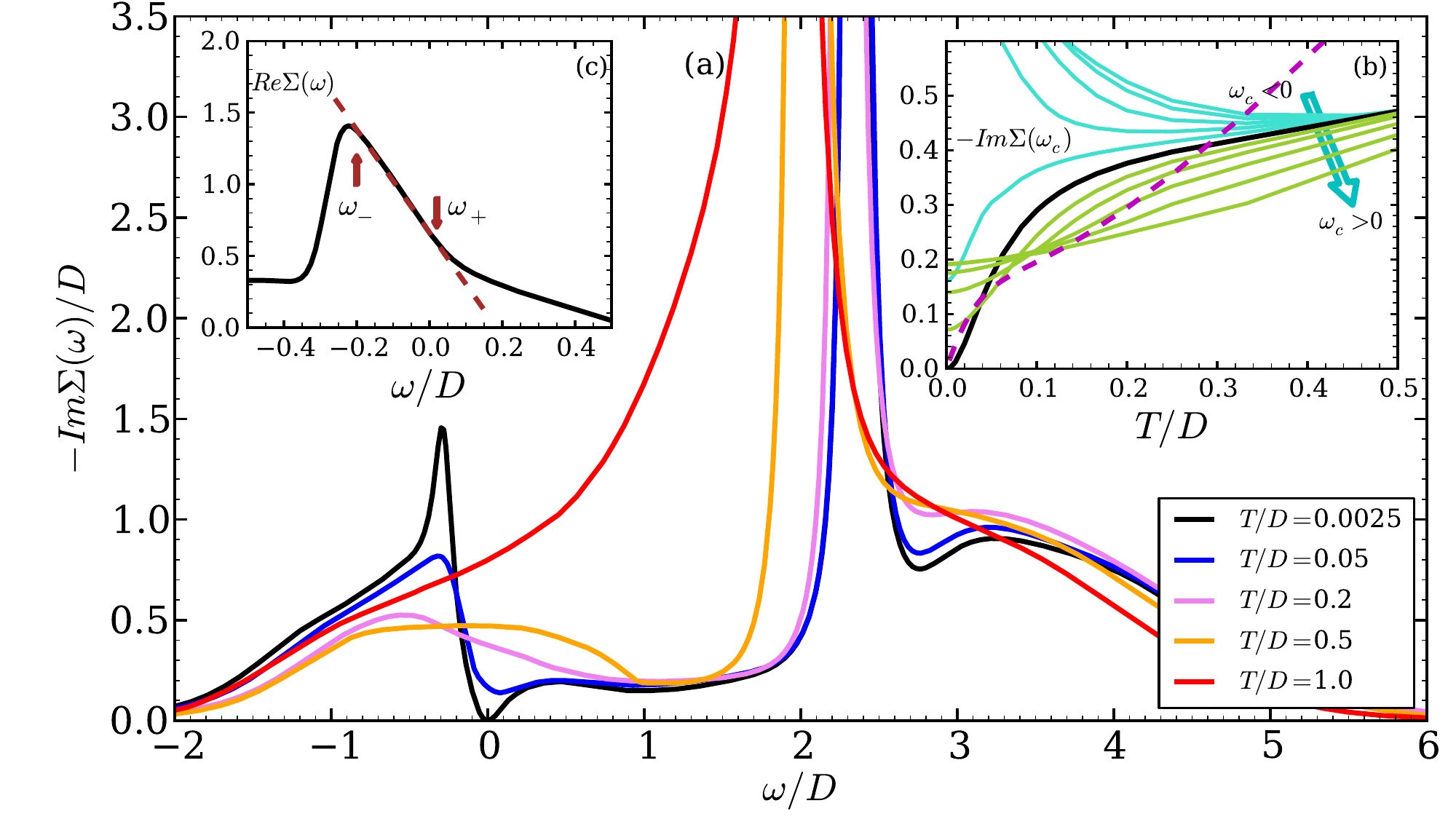}
\caption{Self-energy and particle-hole asymmetry. 
(a) $\mathrm{Im}\Sigma(\omega)$ for different temperatures. 
  (b) Temperature dependence of
  $\mathrm{Im}\Sigma(\omega_c)$ for $\omega_c=-0.5,-0.4,\ldots ,-0.1$
  (turquoise), $\omega_c=0.0$ (thick black) and $\omega_c=0.1,0.2,\ldots ,0.5$
  (green). 
  Below the dashed line $Z(T)\mathrm{Im}\Sigma\lesssim T$.
  (c) $\mathrm{Re}\Sigma(\omega)$ at $T/D=0.0025$, and the two `kinks' (arrows). 
\label{fig:rate}}
\end{figure}
Examination of the self-energy 
(Fig.~\ref{fig:rate}) helps understanding the nature of the QP
excitations, as well as of the different transport regimes.
In local DMFT, $-Z\Imag\Sigma$ can be interpreted at
low-$\omega$ as the inverse of the QP lifetime, and $-\Imag\Sigma$ as the transport scattering rate.
\frate b displays $\Imag\Sigma(\omega,T)$ \vs $T$ for different
excitation energies $\omega$.
At $\omega=0$ (thick curve), FL behavior $\Imag\Sigma
\sim T^2$ applies at low-$T$, corresponding to very long-lived QPs. 
Note that strict FL behavior breaks down already below the temperature at which 
$-Z\mathrm{Im}\Sigma(0,T)\sim T$~\cite{mravlje11}. 
%
%
%
%
%
%
%
At finite frequency, the hole-like excitations have %
higher scattering rate than 
electron-like ones (\frate a,b).
At $\Tknee\simeq 0.08 D$,
the curves of $\isigma$ \vs $T$  
for different positive $\omega$'s display a crossing point. Above $\Tknee$, the scattering rate is a {\it 
decreasing} function of frequency: low-energy electron-like excitations with finite $\omega >0$ have 
a longer lifetime than those at $\omega=0$. 
These finite-energy $\omega>0$ excitations provide the largest contribution to 
conductivity in the intermediate RQP regime~\footnote{For $T\lesssim\TM$, 
the dc conductivity from (\ref{eq:opt_con}) can be approximated by a generalized 
Drude-like formula 
$\sigma_{dc} \propto \int d\omega f'(\omega) \Phi(\epsilon^*_{\omega,T})/\isigma(\omega)$ 
with $\epsilon^*_{\omega,T}\equiv \mu+\omega-\rsigma$. This approximation is highly accurate up to $\Tknee$. The $\omega>0$ 
range provides the largest contribution in the RQP regime.}.
%
%
%
%
%
%
Their inverse lifetime depends weakly on temperature for $T>\Tknee$ (
almost saturated behavior in \frate b ) 
and remains much smaller than the bandwidth and at most comparable to $k_BT$.  
For early considerations on a QP description of transport beyond the FL regime 
in the context of electron-phonon interactions, see~\cite{prange_kadanoff_1964}. 

The strong electron-hole asymmetry has also other interesting consequences.  Because in
the RQP regime %
the $\omega<0$ states are strongly damped, the Fermi surface as determined by the
maximum intensity of $A_{\mathbf k}(\omega=0)$ `inflates' to a %
larger volume than the $T=0$ Luttinger volume~\cite{supp}.  
From \frate c, it is also seen that the deviation from linearity of $\rsigma$ 
($\sim \Sigma_0+\omega(1-1/Z)$ at low $\omega$) defines two distinct 
energy scales, $-\omm$ for hole-like and $\omp$ for electron-like excitations, 
leading to `kinks' in QP dispersions, as documented by previous %
studies~\cite{byczuk_kinks_2007,grete_prb_2011}. 
We note that the smallest kink energy $\omp\ll\omm$ sets the scale for deviations from 
FL behavior ($\omp\simeq \pi\TFL$). 
Using quite different theoretical methods, previous
studies~\cite{zemljic_prelovsek_prl_2008,shastry_prl_2012} have also
emphasized the importance of particle-hole asymmetry in hole-doped
Mott insulators. 

%
%
%
\begin{figure}
\includegraphics[width=1\columnwidth]{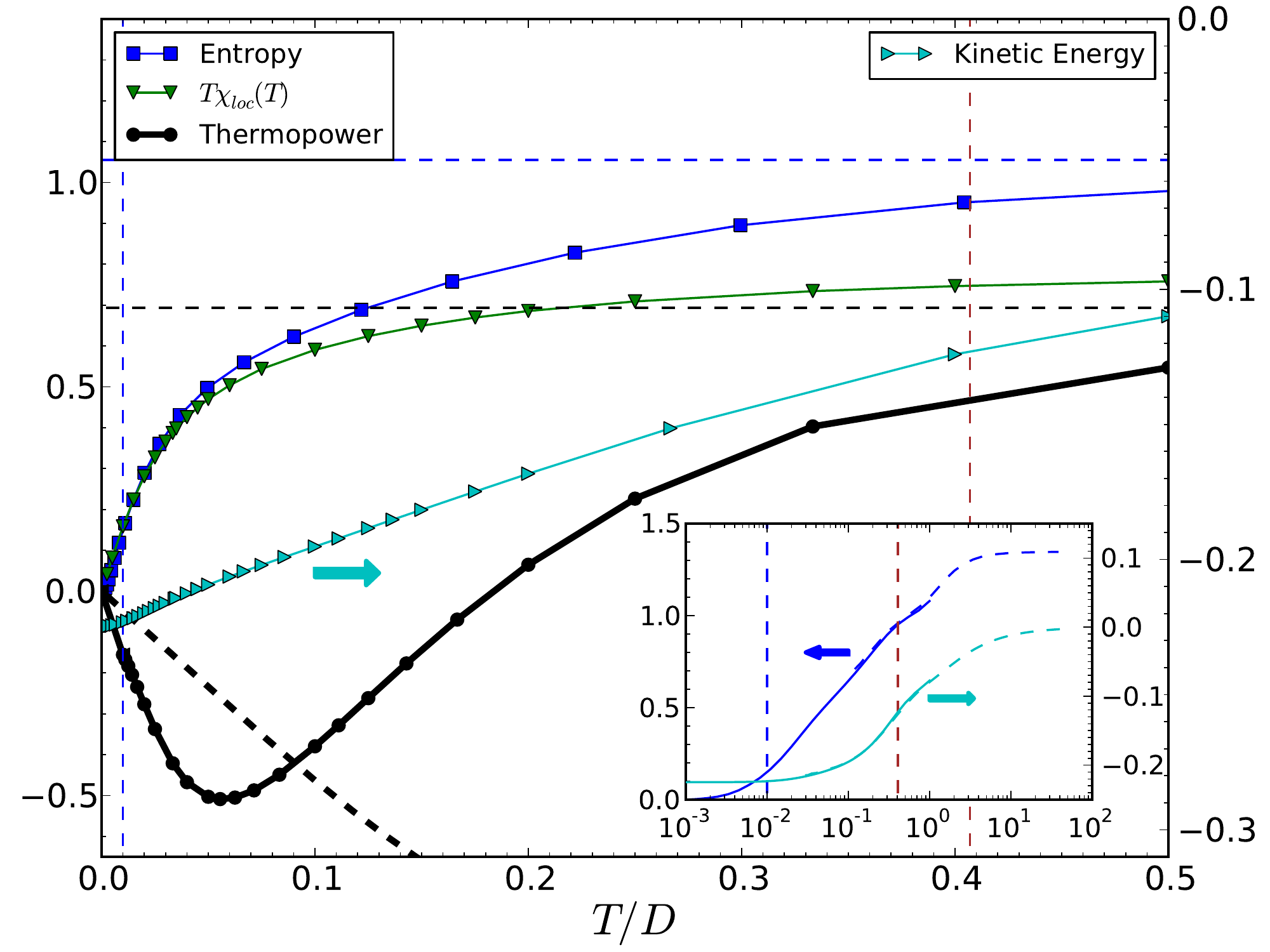}
\caption{Entropy (in units of $k_B$), $T\chi_\mathrm{loc}$
  ($\mu_B=1$), 
  and Seebeck coefficient (in units of $k_B/|e|$), for $\delta=20\%$.  
  Horizontal dashed lines indicate the atomic (Heikes) limit for the entropy and thermopower. 
  Vertical dashed lines denote $\TFL$ and  $\TM$.
The thermopower at low-$T$ is compared to the FL estimate (thick dashed) in which the particle-hole asymmetry 
of $\mathrm{Im}\Sigma$ is neglected. 
Kinetic energy in units of $D$ (right scale). 
Inset: Entropy and kinetic energy versus temperature on a log scale.
\label{fig:thermo}}
\end{figure}
A sensitive probe of the particle-hole asymmetry is the Seebeck coefficient (thermopower) $Q(T)$ shown on \fth.
Strikingly, the subleading particle-hole asymmetric terms in the low-frequency expansion of $\isigma$ 
modify the slope of $Q(T)$ at low-$T$ by a factor of about two, as compared
to a naive FL theory estimate (thick dashed line) that would
only retain terms $\sim\omega^2+(\pi T)^2$. %
This effect was anticipated in Ref.~\cite{Haule_proc09} and is %
shown here to be quantitatively important. 
The 
near saturation of the scattering rate of the RQPs, discussed
above, is also responsible for $Q(T)$ still increasing in an
electron-like manner up to $T\simeq\Tknee$.
At a higher temperature within the %
RQP regime $Q(T)$ changes sign and, %
when entering the bad-metal regime, approaches  
the simple Heikes estimate for $D\lesssim T \ll U$ (\fth). %
The atomic Kelvin formula~\cite{shastry_kelvin_prb_2010,supp}
successfully describes the thermopower there, which can thus be taken
as another fingerprint of a bad-metal. The accuracy of approximate
formulas for thermopower has been tested also in other 
studies~\cite{uchida_2011,wenhu_2011}.

It is interesting to observe how the different transport regimes
relate to thermodynamic observables. On \fth  we display the entropy
$S(T)$, the kinetic energy $K(T)$, and the Curie constant $T\chiloc$
associated with the %
local ($\mathbf{q}$-integrated)
magnetic susceptibility.  The entropy as well as the Curie constant
reach remarkably high values already below $T_\mathrm{MIR}$.  
In the RQP regime, the system thus has to be thought of in
terms of two fluids: a mixture of local moments and of the resilient
QP states. Above $T_\mathrm{MIR}$, the entropic
contribution to the free energy overcomes the kinetic energy gain. The
system is fully incoherent, its entropy approaches the atomic limit and
the Curie constant saturates.

\begin{figure}
\includegraphics[width=1\columnwidth]{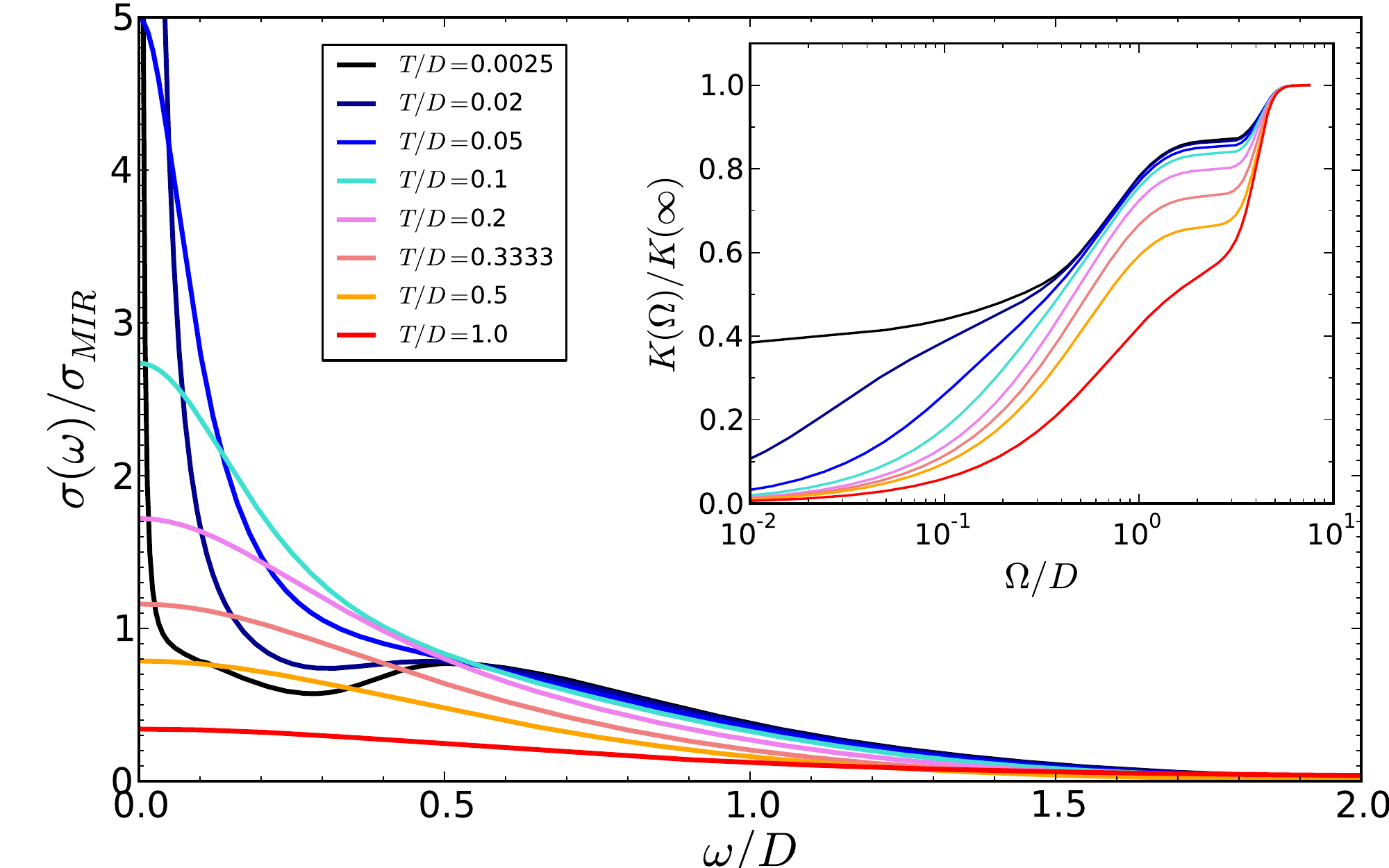}
\caption{Optical conductivity for $\delta=20\%$. (Inset:) Optical spectral
  weight integrated up to $\Omega$, normalized to the kinetic energy.
\label{fig:optics}}
\end{figure}
Optical spectroscopy (\fopt) %
can be used to detect the
two crossovers, between the FL and the RQP regime and from the %
latter into the bad-metal. 
For $T<\TFL$, $\sigma(\omega)$ displays a narrow low-frequency peak
which %
decays as $1/\omega^2$.  This Drude peak corresponds to optical transitions
involving only QP states and has a spectral weight
proportional to doping level $\delta$.
In the FL regime, the Drude peak is well separated from a higher
frequency `hump' (at $\omega/D\sim 0.5$ in \fopt) which corresponds to
transitions between the LHB and QP
states~\cite{jarrell_prb_1995}. The distance of the LHB to the Fermi level, of order $\mu$ (a fraction of the 
bandwidth) sets the energy scale for these transitions. %
 This typically corresponds to the mid-infrared range in narrow-band correlated materials. 
The crossover out of the FL regime into the `resilient metal' regime
leads to a broader low-frequency peak whose frequency dependence is no
longer $1/\omega^2$.  For $T/D=0.02$  $\sigma(\omega)$ can be fit to $1/\omega^{\alpha}$ with
$\alpha\approx 1.2$ for frequencies $0.02 \lesssim \omega/D
\lesssim 0.2$.  The low-$T$ data  display an interesting
`non-Drude foot' with weak frequency dependence of $\sigma(\omega)$ between the Drude and the mid-infrared peaks.
%
%
%
%
%
Above $\Tknee$ the Drude and mid-infrared features merge.

For a rather extended temperature range into the RQP %
regime, the spectral weight redistribution upon increasing $T$ takes
place essentially entirely between the low-energy QP states and the
`mid-infrared' feature, %
see inset of \fopt.
For $T\gtrsim \Tknee$ (warmer part of the intermediate regime), some
spectral weight transfer to higher frequencies starts taking place as
well.

The beginning of the crossover into the bad-metal regime is signaled by two
changes in $\sigma(\omega)$ happening at the same temperature of order 
$\delta D$ (it is shown as the onset of shading in Fig.~\ref{fig:rho}c, which
allows to have a clear view of the region where well-defined quasiparticles
exist). First, the isosbestic crossing point (at $\omega/D\simeq 0.5$) is lost
and the low-frequency non-Drude peak is replaced by a very broad peak, which
corresponds to transitions involving only the LHB. Second, spectral weight is
now redistributed over a considerable energy range, extending all the way to
the UHB (inset).
In the bad-metal, the kinetic energy (to which $\sigma(\omega)$ integrates) is
strongly dependent on temperature (\fth) and the broad peak has correspondingly
a height which continues diminishing with $T$, which implies that the
resistivity does not saturate~\cite{hussey_phil_mag_2004}.

In summary, our study reveals that resilient QP excitations persist well above 
$\TFL$. 
They control transport properties until they disappear at a
temperature roughly of order $\TM$.
%
%
%
The coexistence of QP states with
localized magnetic moments which carry large entropy is intriguing, and
demands closer theoretical investigation including antiferromagnetic
correlations beyond DMFT. %
For hole-doped Mott insulators, a pronounced particle-hole asymmetry is
found.  This calls for new momentum-resolved spectroscopic probes
which would be able to access the `dark side' of the Fermi surface and
for closer investigations of the electron-doped materials where the
signatures of this regime could be seen using conventional ARPES.

\acknowledgements
We would like to thank C.~Berthod, L.~de' Medici, M.~Dressel,
A.~Millis, T.~V.~Ramakrishnan, A.~M.~Tremblay, D.~van der Marel, and
especially N.~Hussey and S.~Shastry for very useful discussions, and
Wenhu Xu for exchanges about thermodynamic observables.  This work was
partially supported by ICAM (X.D.) and the Swiss National Science
Foundation MaNEP program. R.\v{Z}. was supported by ARRS under program
P1-0044.

{\it Note added:} During completion of this manuscript, related results regarding the thermopower 
were independently reported~\cite{canadians_12}.
%


\begin{thebibliography}{10}%
\makeatletter
\providecommand \@ifxundefined [1]{%
 \ifx #1\undefined \expandafter \@firstoftwo
 \else \expandafter \@secondoftwo
\fi
}%
\providecommand \@ifnum [1]{%
 \ifnum #1\expandafter \@firstoftwo
 \else \expandafter \@secondoftwo
\fi
}%
\providecommand \enquote [1]{``#1''}%
\providecommand \bibnamefont  [1]{#1}%
\providecommand \bibfnamefont [1]{#1}%
\providecommand \citenamefont [1]{#1}%
\providecommand\href[0]{\@sanitize\@href}%
\providecommand\@href[1]{\endgroup\@@startlink{#1}\endgroup\@@href}%
\providecommand\@@href[1]{#1\@@endlink}%
\providecommand \@sanitize [0]{\begingroup\catcode`\&12\catcode`\#12\relax}%
\@ifxundefined \pdfoutput {\@firstoftwo}{%
 \@ifnum{\z@=\pdfoutput}{\@firstoftwo}{\@secondoftwo}%
}{%
 \providecommand\@@startlink[1]{\leavevmode}%
 \providecommand\@@endlink[0]{}%
}{%
 \providecommand\@@startlink[1]{%
  \leavevmode
  \pdfstartlink
   attr{/Border[0 0 1 ]/H/I/C[0 1 1]}%
   user{/Subtype/Link/A<</Type/Action/S/URI/URI(#1)>>}%
  \relax
 }%
 \providecommand\@@endlink[0]{\pdfendlink}%
}%
\providecommand \url  [0]{\begingroup\@sanitize \@url }%
\providecommand \@url [1]{\endgroup\@href {#1}{\urlprefix}}%
\providecommand \urlprefix [0]{URL }%
\providecommand \Eprint[0]{\href }%
\@ifxundefined \urlstyle {%
  \providecommand \doi [1]{doi:\discretionary{}{}{}#1}%
}{%
  \providecommand \doi [0]{doi:\discretionary{}{}{}\begingroup
  \urlstyle{rm}\Url }%
}%
\providecommand \doibase [0]{http://dx.doi.org/}%
\providecommand \Doi[1]{\href{\doibase#1}}%
\providecommand \bibAnnote [3]{%
  \BibitemShut{#1}%
  \begin{quotation}\noindent
    \textsc{Key:}\ #2\\\textsc{Annotation:}\ #3%
  \end{quotation}%
}%
\providecommand \bibAnnoteFile [2]{%
  \IfFileExists{#2}{\bibAnnote {#1} {#2} {\input{#2}}}{}%
}%
\providecommand \typeout [0]{\immediate \write \m@ne }%
\providecommand \selectlanguage [0]{\@gobble}%
\providecommand \bibinfo [0]{\@secondoftwo}%
\providecommand \bibfield [0]{\@secondoftwo}%
\providecommand \translation [1]{[#1]}%
\providecommand \BibitemOpen[0]{}%
\providecommand \bibitemStop [0]{}%
\providecommand \bibitemNoStop [0]{.\EOS\space}%
\providecommand \EOS [0]{\spacefactor3000\relax}%
\providecommand \BibitemShut [1]{\csname bibitem#1\endcsname}%
\bibitem{emery_kivelson_prl_1995}%
  \BibitemOpen
  \bibfield{author}{%
  \bibinfo {author} {\bibfnamefont{V.~J.}\ \bibnamefont{Emery}}\ and\ \bibinfo
  {author} {\bibfnamefont{S.~A.}\ \bibnamefont{Kivelson}},\ }%
  \bibfield{journal}{%
  \bibinfo {journal} {Phys. Rev. Lett.}\ }%
  \textbf{\bibinfo {volume} {74}},\ \bibinfo {pages} {3253} (\bibinfo {year}
  {1995})%
  \bibAnnoteFile{NoStop}{emery_kivelson_prl_1995}%
\bibitem{hussey_phil_mag_2004}%
  \BibitemOpen
  \bibfield{author}{%
  \bibinfo {author} {\bibfnamefont{N.}~\bibnamefont{Hussey}}, \bibinfo {author}
  {\bibfnamefont{K.}~\bibnamefont{Takenaka}},\ and\ \bibinfo {author}
  {\bibfnamefont{H.}~\bibnamefont{Takagi}},\ }%
  \bibfield{journal}{%
  \bibinfo {journal} {Phil. Mag.}\ }%
  \textbf{\bibinfo {volume} {84}},\ \bibinfo {pages} {2847} (\bibinfo {year}
  {2004})%
  \bibAnnoteFile{NoStop}{hussey_phil_mag_2004}%
\bibitem{gunnarsson_rmp_2003}%
  \BibitemOpen
  \bibfield{author}{%
  \bibinfo {author} {\bibfnamefont{O.}~\bibnamefont{Gunnarsson}}, \bibinfo
  {author} {\bibfnamefont{M.}~\bibnamefont{Calandra}},\ and\ \bibinfo {author}
  {\bibfnamefont{J.~E.}\ \bibnamefont{Han}},\ }%
  \bibfield{journal}{%
  \bibinfo {journal} {Rev. Mod. Phys.}\ }%
  \textbf{\bibinfo {volume} {75}},\ \bibinfo {pages} {1085} (\bibinfo {year}
  {2003})%
  \bibAnnoteFile{NoStop}{gunnarsson_rmp_2003}%
\bibitem{merino_mckenzie_prb_2000}%
  \BibitemOpen
  \bibfield{author}{%
  \bibinfo {author} {\bibfnamefont{J.}~\bibnamefont{Merino}}\ and\ \bibinfo
  {author} {\bibfnamefont{R.~H.}\ \bibnamefont{McKenzie}},\ }%
  \bibfield{journal}{%
  \bibinfo {journal} {Phys. Rev. B}\ }%
  \textbf{\bibinfo {volume} {61}},\ \bibinfo {pages} {7996} (\bibinfo {year}
  {2000})%
  \bibAnnoteFile{NoStop}{merino_mckenzie_prb_2000}%
\bibitem{limelette_prl_2003}%
  \BibitemOpen
  \bibfield{author}{%
  \bibinfo {author} {\bibfnamefont{P.}~\bibnamefont{Limelette}}, \bibinfo
  {author} {\bibfnamefont{P.}~\bibnamefont{Wzietek}}, \bibinfo {author}
  {\bibfnamefont{S.}~\bibnamefont{Florens}}, \bibinfo {author}
  {\bibfnamefont{A.}~\bibnamefont{Georges}}, \bibinfo {author}
  {\bibfnamefont{T.~A.}\ \bibnamefont{Costi}}, \bibinfo {author}
  {\bibfnamefont{C.}~\bibnamefont{Pasquier}}, \bibinfo {author}
  {\bibfnamefont{D.}~\bibnamefont{J\'erome}}, \bibinfo {author}
  {\bibfnamefont{C.}~\bibnamefont{M\'ezi\`ere}},\ and\ \bibinfo {author}
  {\bibfnamefont{P.}~\bibnamefont{Batail}},\ }%
  \bibfield{journal}{%
  \bibinfo {journal} {Phys. Rev. Lett.}\ }%
  \textbf{\bibinfo {volume} {91}},\ \bibinfo {pages} {016401} (\bibinfo {year}
  {2003})%
  \bibAnnoteFile{NoStop}{limelette_prl_2003}%
\bibitem{merino_prl_2008}%
  \BibitemOpen
  \bibfield{author}{%
  \bibinfo {author} {\bibfnamefont{J.}~\bibnamefont{Merino}}, \bibinfo {author}
  {\bibfnamefont{M.}~\bibnamefont{Dumm}}, \bibinfo {author}
  {\bibfnamefont{N.}~\bibnamefont{Drichko}}, \bibinfo {author}
  {\bibfnamefont{M.}~\bibnamefont{Dressel}},\ and\ \bibinfo {author}
  {\bibfnamefont{R.~H.}\ \bibnamefont{McKenzie}},\ }%
  \bibfield{journal}{%
  \bibinfo {journal} {Phys. Rev. Lett.}\ }%
  \textbf{\bibinfo {volume} {100}},\ \bibinfo {pages} {086404} (\bibinfo {year}
  {2008})%
  \bibAnnoteFile{NoStop}{merino_prl_2008}%
\bibitem{Hussey_Nature_2003}%
  \BibitemOpen
  \bibfield{author}{%
  \bibinfo {author} {\bibfnamefont{N.~E.}\ \bibnamefont{Hussey}}, \bibinfo
  {author} {\bibfnamefont{M.}~\bibnamefont{Abdel-Jawad}}, \bibinfo {author}
  {\bibfnamefont{A.}~\bibnamefont{Carrington}}, \bibinfo {author}
  {\bibfnamefont{A.~P.}\ \bibnamefont{Mackenzie}},\ and\ \bibinfo {author}
  {\bibfnamefont{L.}~\bibnamefont{Balicas}},\ }%
  \bibfield{journal}{%
  \Doi{10.1038/nature01981}{\bibinfo {journal} {Nature}}\ }%
  \textbf{\bibinfo {volume} {425}},\ \bibinfo {pages} {814} (\bibinfo {year}
  {2003})%
  \bibAnnoteFile{NoStop}{Hussey_Nature_2003}%
\bibitem{doiron_proust_nature_2007}%
  \BibitemOpen
  \bibfield{author}{%
  \bibinfo {author} {\bibfnamefont{N.}~\bibnamefont{Doiron-Leyraud}}, \bibinfo
  {author} {\bibfnamefont{C.}~\bibnamefont{Proust}}, \bibinfo {author}
  {\bibfnamefont{D.}~\bibnamefont{LeBoeuf}}, \bibinfo {author}
  {\bibfnamefont{J.}~\bibnamefont{Levallois}}, \bibinfo {author}
  {\bibfnamefont{J.-B.}\ \bibnamefont{Bonnemaison}}, \bibinfo {author}
  {\bibfnamefont{R.}~\bibnamefont{Liang}}, \bibinfo {author}
  {\bibfnamefont{D.~A.}\ \bibnamefont{Bonn}}, \bibinfo {author}
  {\bibfnamefont{W.~N.}\ \bibnamefont{Hardy}},\ and\ \bibinfo {author}
  {\bibfnamefont{L.}~\bibnamefont{Taillefer}},\ }%
  \bibfield{journal}{%
  \bibinfo {journal} {Nature}\ }%
  \textbf{\bibinfo {volume} {447}},\ \bibinfo {pages} {565} (\bibinfo {year}
  {2007})%
  \bibAnnoteFile{NoStop}{doiron_proust_nature_2007}%
\bibitem{hussey_jphys_2008}%
  \BibitemOpen
  \bibfield{author}{%
  \bibinfo {author} {\bibfnamefont{N.~E.}\ \bibnamefont{Hussey}},\ }%
  \bibfield{journal}{%
  \bibinfo {journal} {Journal of Physics: Condensed Matter}\ }%
  \textbf{\bibinfo {volume} {20}},\ \bibinfo {pages} {123201}%
  \bibAnnoteFile{NoStop}{hussey_jphys_2008}%
\bibitem{barisic_12}%
  \BibitemOpen
  \bibfield{author}{%
  \bibinfo {author} {\bibfnamefont{M.}~\bibnamefont{Greven}}\ and\ \bibinfo
  {author} {\bibfnamefont{N.}~\bibnamefont{Bari\v{s}i\'c}}}%
   (\bibinfo {year} {2012}),\ \bibinfo {note} {internal communication}%
  \bibAnnoteFile{NoStop}{barisic_12}%
\bibitem{mirzaei_vdMarel_2012}%
  \BibitemOpen
  \bibfield{author}{%
  \bibinfo {author} {\bibfnamefont{S.~I.}\ \bibnamefont{Mirzaei}}, \bibinfo
  {author} {\bibfnamefont{D.}~\bibnamefont{Stricker}}, \bibinfo {author}
  {\bibfnamefont{J.~N.}\ \bibnamefont{Hancock}}, \bibinfo {author}
  {\bibfnamefont{C.}~\bibnamefont{Berthod}}, \bibinfo {author}
  {\bibfnamefont{A.}~\bibnamefont{Georges}}, \bibinfo {author}
  {\bibfnamefont{E.}~\bibnamefont{van Heumen}}, \bibinfo {author}
  {\bibfnamefont{M.~K.}\ \bibnamefont{Chan}}, \bibinfo {author}
  {\bibfnamefont{X.}~\bibnamefont{Zhao}}, \bibinfo {author}
  {\bibfnamefont{Y.}~\bibnamefont{Li}}, \bibinfo {author}
  {\bibfnamefont{M.}~\bibnamefont{Greven}}, \bibinfo {author}
  {\bibfnamefont{N.}~\bibnamefont{Bari\v{s}i\'c}},\ and\ \bibinfo {author}
  {\bibfnamefont{D.}~\bibnamefont{van~der Marel}},\ }%
  \enquote{\bibinfo {title} {Evidence for a fermi liquid in the pseudogap phase
  of high-tc cuprates},}\ \bibinfo {note} {ArXiv:1207.6704}%
  \bibAnnoteFile{NoStop}{mirzaei_vdMarel_2012}%
\bibitem{pruschke93}%
  \BibitemOpen
  \bibfield{author}{%
  \bibinfo {author} {\bibfnamefont{T.}~\bibnamefont{Pruschke}}, \bibinfo
  {author} {\bibfnamefont{D.}~\bibnamefont{Cox}},\ and\ \bibinfo {author}
  {\bibfnamefont{M.}~\bibnamefont{Jarrell}},\ }%
  \bibfield{journal}{%
  \bibinfo {journal} {Europhys. Lett.}\ }%
  \textbf{\bibinfo {volume} {21}},\ \bibinfo {pages} {593} (\bibinfo {year}
  {1993})%
  \bibAnnoteFile{NoStop}{pruschke93}%
\bibitem{jarrell_prb_1994}%
  \BibitemOpen
  \bibfield{author}{%
  \bibinfo {author} {\bibfnamefont{M.}~\bibnamefont{Jarrell}}\ and\ \bibinfo
  {author} {\bibfnamefont{T.}~\bibnamefont{Pruschke}},\ }%
  \bibfield{journal}{%
  \bibinfo {journal} {Phys. Rev. B}\ }%
  \textbf{\bibinfo {volume} {49}},\ \bibinfo {pages} {1458} (\bibinfo {year}
  {1994})%
  \bibAnnoteFile{NoStop}{jarrell_prb_1994}%
\bibitem{kajueter_prb_1996}%
  \BibitemOpen
  \bibfield{author}{%
  \bibinfo {author} {\bibfnamefont{H.}~\bibnamefont{Kajueter}}, \bibinfo
  {author} {\bibfnamefont{G.}~\bibnamefont{Kotliar}},\ and\ \bibinfo {author}
  {\bibfnamefont{G.}~\bibnamefont{Moeller}},\ }%
  \bibfield{journal}{%
  \bibinfo {journal} {Phys. Rev. B}\ }%
  \textbf{\bibinfo {volume} {53}},\ \bibinfo {pages} {16214} (\bibinfo {year}
  {1996})%
  \bibAnnoteFile{NoStop}{kajueter_prb_1996}%
\bibitem{palsson_prl_1998}%
  \BibitemOpen
  \bibfield{author}{%
  \bibinfo {author} {\bibfnamefont{G.}~\bibnamefont{P\'alsson}}\ and\ \bibinfo
  {author} {\bibfnamefont{G.}~\bibnamefont{Kotliar}},\ }%
  \bibfield{journal}{%
  \bibinfo {journal} {Phys. Rev. Lett.}\ }%
  \textbf{\bibinfo {volume} {80}},\ \bibinfo {pages} {4775} (\bibinfo {year}
  {1998})%
  \bibAnnoteFile{NoStop}{palsson_prl_1998}%
\bibitem{palsson_thesis_2001}%
  \BibitemOpen
  \bibfield{author}{%
  \bibinfo {author} {\bibfnamefont{G.}~\bibnamefont{Palsson}},\ }%
  \emph{\bibinfo {title} {Computational studies of thermoelectricity in
  strongly correlated electron systems}},\ Ph.D. thesis,\ \bibinfo {school}
  {Rutgers University, NJ} (\bibinfo {year} {2001})%
  \bibAnnoteFile{NoStop}{palsson_thesis_2001}%
\bibitem{jarrell_prb_1995}%
  \BibitemOpen
  \bibfield{author}{%
  \bibinfo {author} {\bibfnamefont{M.}~\bibnamefont{Jarrell}}, \bibinfo
  {author} {\bibfnamefont{J.~K.}\ \bibnamefont{Freericks}},\ and\ \bibinfo
  {author} {\bibfnamefont{T.}~\bibnamefont{Pruschke}},\ }%
  \bibfield{journal}{%
  \bibinfo {journal} {Phys. Rev. B}\ }%
  \textbf{\bibinfo {volume} {51}},\ \bibinfo {pages} {11704} (\bibinfo {year}
  {1995})%
  \bibAnnoteFile{NoStop}{jarrell_prb_1995}%
\bibitem{georges_rmp_1996}%
  \BibitemOpen
  \bibfield{author}{%
  \bibinfo {author} {\bibfnamefont{A.}~\bibnamefont{Georges}}, \bibinfo
  {author} {\bibfnamefont{G.}~\bibnamefont{Kotliar}}, \bibinfo {author}
  {\bibfnamefont{W.}~\bibnamefont{Krauth}},\ and\ \bibinfo {author}
  {\bibfnamefont{M.~J.}\ \bibnamefont{Rozenberg}},\ }%
  \bibfield{journal}{%
  \bibinfo {journal} {Rev. Mod. Phys.}\ }%
  \textbf{\bibinfo {volume} {68}},\ \bibinfo {pages} {13} (\bibinfo {year}
  {1996})%
  \bibAnnoteFile{NoStop}{georges_rmp_1996}%
\bibitem{bulla_rmp_2008}%
  \BibitemOpen
  \bibfield{author}{%
  \bibinfo {author} {\bibfnamefont{R.}~\bibnamefont{Bulla}}, \bibinfo {author}
  {\bibfnamefont{T.~A.}\ \bibnamefont{Costi}},\ and\ \bibinfo {author}
  {\bibfnamefont{T.}~\bibnamefont{Pruschke}},\ }%
  \bibfield{journal}{%
  \bibinfo {journal} {Rev. Mod. Phys.}\ }%
  \textbf{\bibinfo {volume} {80}},\ \bibinfo {pages} {395} (\bibinfo {year}
  {2008})%
  \bibAnnoteFile{NoStop}{bulla_rmp_2008}%
\bibitem{NRGLjubljana}%
  \BibitemOpen
  \bibfield{author}{%
  \bibinfo {author} {\bibfnamefont{R.}~\bibnamefont{\v{Z}itko}},\ }%
  \enquote{\bibinfo {title} {"NRG ljubljana" - open source numerical
  renormalization group code},}\ \url{http://nrgljubljana.ijs.si/}%
  \bibAnnoteFile{NoStop}{NRGLjubljana}%
\bibitem{gull_rmp_2011}%
  \BibitemOpen
  \bibfield{author}{%
  \bibinfo {author} {\bibfnamefont{E.}~\bibnamefont{Gull}}, \bibinfo {author}
  {\bibfnamefont{A.~J.}\ \bibnamefont{Millis}}, \bibinfo {author}
  {\bibfnamefont{A.~I.}\ \bibnamefont{Lichtenstein}}, \bibinfo {author}
  {\bibfnamefont{A.~N.}\ \bibnamefont{Rubtsov}}, \bibinfo {author}
  {\bibfnamefont{M.}~\bibnamefont{Troyer}},\ and\ \bibinfo {author}
  {\bibfnamefont{P.}~\bibnamefont{Werner}},\ }%
  \bibfield{journal}{%
  \bibinfo {journal} {Rev. Mod. Phys.}\ }%
  \textbf{\bibinfo {volume} {83}},\ \bibinfo {pages} {349} (\bibinfo {year}
  {2011})%
  \bibAnnoteFile{NoStop}{gull_rmp_2011}%
\bibitem{TRIQS}%
  \BibitemOpen
  \bibfield{author}{%
  \bibinfo {author} {\bibfnamefont{M.}~\bibnamefont{Ferrero}}\ and\ \bibinfo
  {author} {\bibfnamefont{O.}~\bibnamefont{Parcollet}},\ }%
  \enquote{\bibinfo {title} {{TRIQS}: a {T}oolbox for {R}esearch on
  {I}nteracting {Q}uantum {S}ystems},}\ \url{http://ipht.cea.fr/triqs}%
  \bibAnnoteFile{NoStop}{TRIQS}%
\bibitem{supp}%
  \BibitemOpen
  \enquote{\bibinfo {title} {Online supplementary material},}\ %
  \bibAnnoteFile{NoStop}{supp}%
\bibitem{mravlje11}%
  \BibitemOpen
  \bibfield{author}{%
  \bibinfo {author} {\bibfnamefont{J.}~\bibnamefont{Mravlje}}, \bibinfo
  {author} {\bibfnamefont{M.}~\bibnamefont{Aichhorn}}, \bibinfo {author}
  {\bibfnamefont{T.}~\bibnamefont{Miyake}}, \bibinfo {author}
  {\bibfnamefont{K.}~\bibnamefont{Haule}}, \bibinfo {author}
  {\bibfnamefont{G.}~\bibnamefont{Kotliar}},\ and\ \bibinfo {author}
  {\bibfnamefont{A.}~\bibnamefont{Georges}},\ }%
  \bibfield{journal}{%
  \bibinfo {journal} {Phys. Rev. Lett}\ }%
  \textbf{\bibinfo {volume} {106}},\ \bibinfo {pages} {096401} (\bibinfo {year}
  {2011})%
  \bibAnnoteFile{NoStop}{mravlje11}%
\bibitem{Note1}%
  \BibitemOpen
  \bibinfo {note} {For $T\lesssim T_{\protect \mathrm {MIR}}$, the dc
  conductivity from (\ref {eq:opt_con}) can be approximated by a generalized
  Drude-like formula $\sigma _{dc} \propto \DOTSI \intop \ilimits@ d\omega
  f'(\omega ) \Phi (\epsilon ^*_{\omega ,T})/\protect \mathrm {Im}\Sigma
  (\omega )$ with $\epsilon ^*_{\omega ,T}\equiv \mu +\omega -\protect \mathrm
  {Re}\Sigma $. This approximation is highly accurate up to $T_{*}$. The
  $\omega >0$ range provides the largest contribution in the RQP regime.}%
  \bibAnnoteFile{Stop}{Note1}%
\bibitem{prange_kadanoff_1964}%
  \BibitemOpen
  \bibfield{author}{%
  \bibinfo {author} {\bibfnamefont{R.~E.}\ \bibnamefont{Prange}}\ and\ \bibinfo
  {author} {\bibfnamefont{L.~P.}\ \bibnamefont{Kadanoff}},\ }%
  \bibfield{journal}{%
  \bibinfo {journal} {Phys. Rev.}\ }%
  \textbf{\bibinfo {volume} {134}},\ \bibinfo {pages} {A566} (\bibinfo {year}
  {1964})%
  \bibAnnoteFile{NoStop}{prange_kadanoff_1964}%
\bibitem{byczuk_kinks_2007}%
  \BibitemOpen
  \bibfield{author}{%
  \bibinfo {author} {\bibfnamefont{K.}~\bibnamefont{Byczuk}}, \bibinfo {author}
  {\bibfnamefont{M.}~\bibnamefont{Kollar}}, \bibinfo {author}
  {\bibfnamefont{K.}~\bibnamefont{Held}}, \bibinfo {author}
  {\bibfnamefont{Y.-F.}\ \bibnamefont{Yang}}, \bibinfo {author}
  {\bibfnamefont{I.~A.}\ \bibnamefont{Nekrasov}}, \bibinfo {author}
  {\bibfnamefont{T.}~\bibnamefont{Pruschke}},\ and\ \bibinfo {author}
  {\bibfnamefont{D.}~\bibnamefont{Vollhardt}},\ }%
  \bibfield{journal}{%
  \bibinfo {journal} {Nat. Phys.}\ }%
  \textbf{\bibinfo {volume} {3}},\ \bibinfo {pages} {168} (\bibinfo {year} {2007})%
  \bibAnnoteFile{NoStop}{byczuk_kinks_2007}%
\bibitem{grete_prb_2011}%
  \BibitemOpen
  \bibfield{author}{%
  \bibinfo {author} {\bibfnamefont{P.}~\bibnamefont{Grete}}, \bibinfo {author}
  {\bibfnamefont{S.}~\bibnamefont{Schmitt}}, \bibinfo {author}
  {\bibfnamefont{C.}~\bibnamefont{Raas}}, \bibinfo {author}
  {\bibfnamefont{F.~B.}\ \bibnamefont{Anders}},\ and\ \bibinfo {author}
  {\bibfnamefont{G.~S.}\ \bibnamefont{Uhrig}},\ }%
  \bibfield{journal}{%
  \bibinfo {journal} {Phys. Rev. B}\ }%
  \textbf{\bibinfo {volume} {84}},\ \bibinfo {pages} {205104} (\bibinfo {year}
  {2011})%
  \bibAnnoteFile{NoStop}{grete_prb_2011}%
\bibitem{zemljic_prelovsek_prl_2008}%
  \BibitemOpen
  \bibfield{author}{%
  \bibinfo {author} {\bibfnamefont{M.~M.}\
  \bibnamefont{Zemlji\ifmmode~\check{c}\else \v{c}\fi{}}}, \bibinfo {author}
  {\bibfnamefont{P.}~\bibnamefont{Prelov\ifmmode~\check{s}\else
  \v{s}\fi{}ek}},\ and\ \bibinfo {author}
  {\bibfnamefont{T.}~\bibnamefont{Tohyama}},\ }%
  \bibfield{journal}{%
  \bibinfo {journal} {Phys. Rev. Lett.}\ }%
  \textbf{\bibinfo {volume} {100}},\ \bibinfo {pages} {036402} (\bibinfo {year}
  {2008})%
  \bibAnnoteFile{NoStop}{zemljic_prelovsek_prl_2008}%
\bibitem{shastry_prl_2012}%
  \BibitemOpen
  \bibfield{author}{%
  \bibinfo {author} {\bibfnamefont{B.~S.}\ \bibnamefont{Shastry}},\ }%
  \bibfield{journal}{%
  \bibinfo {journal} {Phys. Rev. Lett.}\ }%
  \textbf{\bibinfo {volume} {109}},\ \bibinfo {pages} {067004} (\bibinfo {year}
  {2012})%
  \bibAnnoteFile{NoStop}{shastry_prl_2012}%
\bibitem{Haule_proc09}%
  \BibitemOpen
  \bibfield{author}{%
  \bibinfo {author} {\bibfnamefont{K.}~\bibnamefont{Haule}}\ and\ \bibinfo
  {author} {\bibfnamefont{G.}~\bibnamefont{Kotliar}},\ }%
  in\ \emph{\bibinfo {booktitle} {Properties and Applications of Thermoelectric
  Materials}},\ \bibinfo {series and number} {NATO Science for Peace and
  Security Series B: Physics and Biophysics},\ \bibinfo {editor} {edited by\
  \bibinfo {editor} {\bibfnamefont{V.}~\bibnamefont{Zlati\'c}}\ and\ \bibinfo
  {editor} {\bibfnamefont{A.~C.}\ \bibnamefont{Hewson}}}\ (\bibinfo {publisher}
  {Springer Netherlands},\ \bibinfo {year} {2009})\ pp.\ \bibinfo {pages}
  {119--131}%
  \bibAnnoteFile{NoStop}{Haule_proc09}%
\bibitem{shastry_kelvin_prb_2010}%
  \BibitemOpen
  \bibfield{author}{%
  \bibinfo {author} {\bibfnamefont{M.~R.}\ \bibnamefont{Peterson}}\ and\
  \bibinfo {author} {\bibfnamefont{B.~S.}\ \bibnamefont{Shastry}},\ }%
  \bibfield{journal}{%
  \bibinfo {journal} {Phys. Rev. B}\ }%
  \textbf{\bibinfo {volume} {82}},\ \bibinfo {pages} {195105} (\bibinfo {year}
  {2010})%
  \bibAnnoteFile{NoStop}{shastry_kelvin_prb_2010}%
\bibitem{uchida_2011}%
  \BibitemOpen
  \bibfield{author}{%
  \bibinfo {author} {\bibfnamefont{M.}~\bibnamefont{Uchida}}, \bibinfo {author}
  {\bibfnamefont{K.}~\bibnamefont{Oishi}}, \bibinfo {author}
  {\bibfnamefont{M.}~\bibnamefont{Matsuo}}, \bibinfo {author}
  {\bibfnamefont{W.}~\bibnamefont{Koshibae}}, \bibinfo {author}
  {\bibfnamefont{Y.}~\bibnamefont{Onose}}, \bibinfo {author}
  {\bibfnamefont{M.}~\bibnamefont{Mori}}, \bibinfo {author}
  {\bibfnamefont{J.}~\bibnamefont{Fujioka}}, \bibinfo {author}
  {\bibfnamefont{S.}~\bibnamefont{Miyasaka}}, \bibinfo {author}
  {\bibfnamefont{S.}~\bibnamefont{Maekawa}},\ and\ \bibinfo {author}
  {\bibfnamefont{Y.}~\bibnamefont{Tokura}},\ }%
  \bibfield{journal}{%
  \bibinfo {journal} {Phys. Rev. B}\ }%
  \textbf{\bibinfo {volume} {83}},\ \bibinfo {pages} {165127} (\bibinfo {year}
  {2011})%
  \bibAnnoteFile{NoStop}{uchida_2011}%
\bibitem{wenhu_2011}%
  \BibitemOpen
  \bibfield{author}{%
  \bibinfo {author} {\bibfnamefont{W.}~\bibnamefont{Xu}}, \bibinfo {author}
  {\bibfnamefont{C.}~\bibnamefont{Weber}},\ and\ \bibinfo {author}
  {\bibfnamefont{G.}~\bibnamefont{Kotliar}},\ }%
  \bibfield{journal}{%
  \bibinfo {journal} {Phys. Rev. B}\ }%
  \textbf{\bibinfo {volume} {84}},\ \bibinfo {pages} {035114} (\bibinfo {year}
  {2011})%
  \bibAnnoteFile{NoStop}{wenhu_2011}%
\bibitem{canadians_12}%
  \BibitemOpen
  \bibfield{author}{%
  \bibinfo {author} {\bibfnamefont{L.-F.}\ \bibnamefont{Arsenault}}, \bibinfo
  {author} {\bibfnamefont{B.~S.}\ \bibnamefont{Shastry}}, \bibinfo {author}
  {\bibfnamefont{P.}~\bibnamefont{S\'emon}},\ and\ \bibinfo {author}
  {\bibfnamefont{A.-M.~S.}\ \bibnamefont{Tremblay}},\ }%
  \enquote{\bibinfo {title} {Entropy, frustration and large thermopower of
  doped mott insulators on the fcc lattice},}\ \bibinfo {note}
  {ArXiv:1209.4349}%
  \bibAnnoteFile{NoStop}{canadians_12}%
\end{thebibliography}
\newpage
\appendix*
{\begin{center} {\bf \Large Supplementary information}\end{center}}

\section{Self-energy scaling}
\begin{figure}
\includegraphics[width=1\columnwidth]{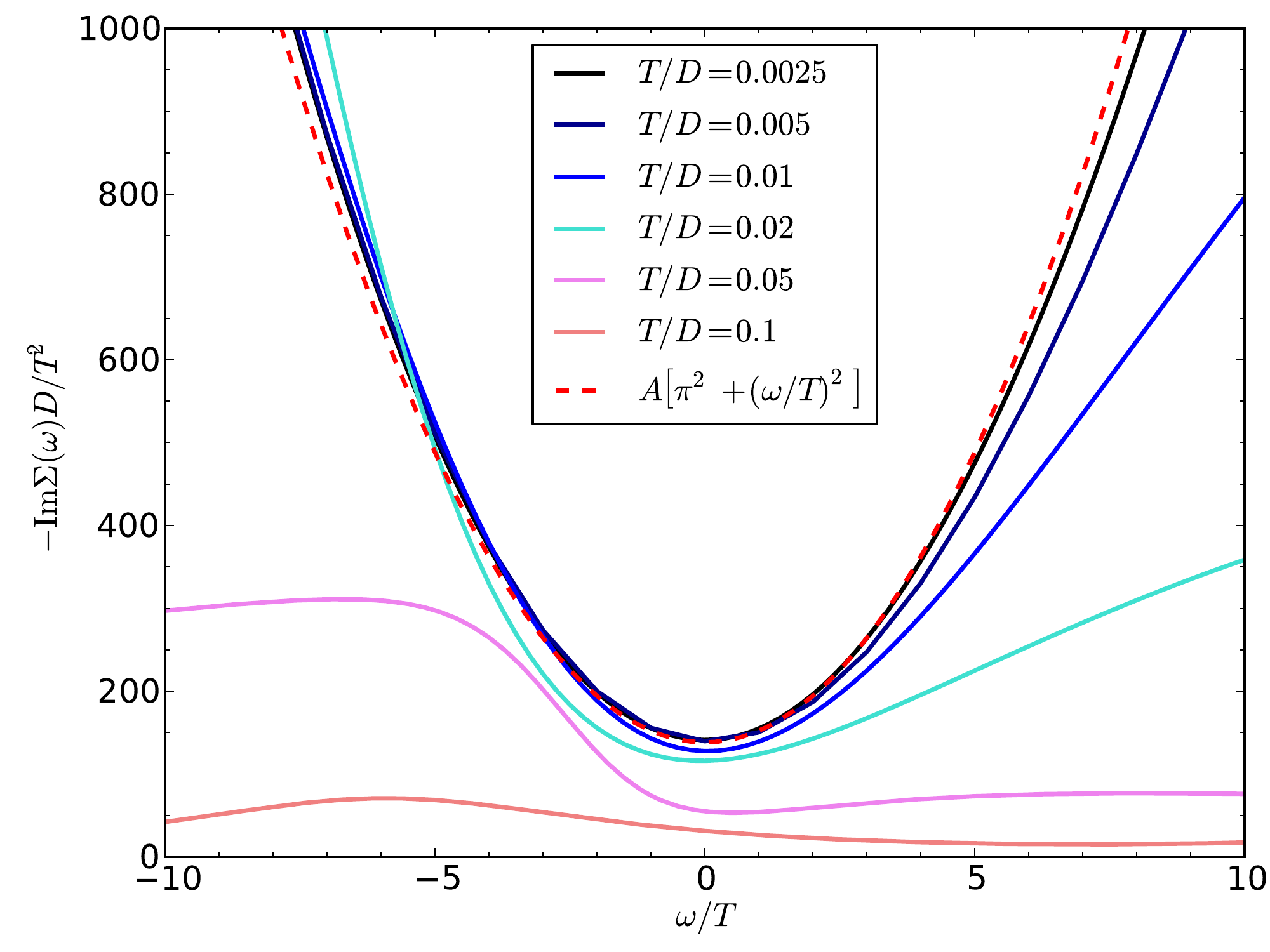}
  \caption{The self energy scaling as $\omega/T$ for the doped Hubbard
    model considered in the main text (Bethe lattice, $U=4D$, doping
    $\delta=0.2$). At low temperatures self energy follows
    Eq.~\ref{eq:scaling}.  $A$ is found to be close to $Z^{-2}\sim
    \delta^{-2}$. (For doping 0.2, $Z=0.22$.)
  \label{fig:FLscale}}
\end{figure}

In a Fermi liquid, $\mathrm{Im}\Sigma(\omega,T) \propto
\left[\omega^2+(\pi T)^2\right]$ and therefore the self-energy obeys
the following scaling in $\omega/T$:
\begin{equation}
-\frac{D\mathrm{Im} \Sigma}{T^2} = A \left[\pi^2+ (\omega/T)^2\right]. \label{eq:scaling}
\end{equation}
In Fig.~\ref{fig:FLscale}, we show $-\mathrm{Im}\Sigma(\omega,T)/T^2$ as a
function of $\omega/T$ for different temperatures. As the temperature is
lowered below $T_\mathrm{FL}\approx0.01D$, the curves collapse on a parabola,
confirming the expected Fermi-liquid scaling law. Finding such a scaling is
actually a stringent test on the numerical data.  Very precise quantum Monte
Carlo data (analytically continued with Pad\'e approximants) and a strict
control of the chemical potential were needed to obtain these results.

As the temperature is raised, the positive-frequency side quickly
deviates from the scaling form, revealing a particle-hole asymmetry
already for temperatures above $T/D=0.005$.  The deviations appear at
a scale $\omega_+ \approx \pi T_{\mathrm{FL}}$. On the
negative-frequency side the self-energy follows the quadratic behavior
much more robustly and deviates from the scaling function only at
about $T/D = 0.02$. Notice that the corrections on the $\omega>0$ side
grow linearly with temperature. This leads to a Seebeck coefficient
which is linear in $T$ at low temperatures, but with an enhanced slope
as compared to the result one would get if these corrections were
neglected.

The transport probes an energy window of a few (say from -5 to 5) $k_B
T$ . In this energy window the self-energy starts to deviate
appreciably at $T/D = 0.01$. This is where the resistivity (within the
precision of our data) visually departs from the $T^2$ law.

\section{Temperature evolution of momentum-resolved spectra}
\begin{figure*}
\includegraphics[width=2\columnwidth]{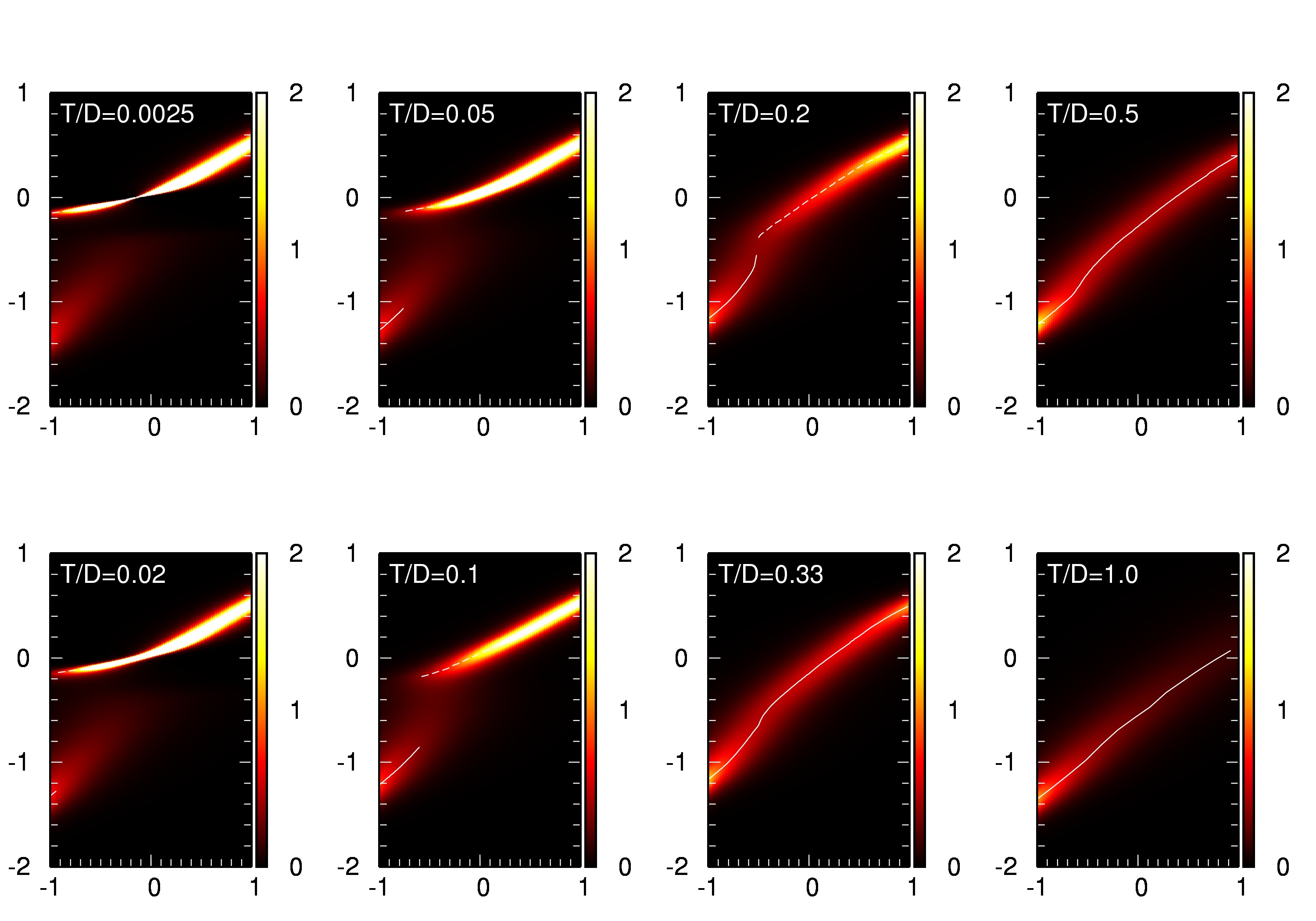}
  \caption{Contour map of momentum-resolved spectra
$A(\epsilon_k,\omega)$ at various temperatures for the same parameters as
the data in the main text: doping $\delta=0.2$ and $U/D=4.0$. \label{fig:contour}}
\end{figure*}

In Fig.~\ref{fig:contour} we plot the temperature evolution of the
momentum-resolved spectra using a color-map where bright (dark) colors
indicate high (low) values of $A_k(\omega)$. 

At low temperatures, the data display two peaks corresponding to: the
lower Hubbard band (LHB) which disperses around
$\omega_{\mathrm{LHB}}\sim -\mu_0$ and the quasiparticle peak (QP) in
the vicinity of $\omega \sim 0$. $\mu_0$ is the 
effective chemical potential at $T=0$. The upper Hubbard band (UHB) centered at the
energy $\omega_{\mathrm{UHB}}\sim U-\mu_0$ lies above the energy range
displayed in the plot.

The lowest temperature data $T/D=0.0025$ show a very sharp QP peak
around $\omega=0$, which is rapidly broadened as the frequency is
increased. At very small frequencies, the slope of the dispersion is
found to be approximately 5 times smaller than the bare one, as
dictated by $Z\approx 0.2$. On the negative-frequency side this holds
almost until the bottom of the band. On the positive-frequency side,
instead, the kink at $\omega_+$ is rapidly encountered. Above this
kink, the slope of the band dispersion increases to about half the
bare dispersion slope. The LHB, seen clearly for occupied states $k<k_F$, disperses
at a slope close to that of the bare dispersion.

As the temperature is increased, the QP band broadens and becomes more
dispersing. It becomes therefore progressively more difficult to
resolve it from the LHB band. Nevertheless, for temperatures well
above $T_\mathrm{FL}$ and $T_*$ (four leftmost panels) one can still
clearly distinguish the QP band from the LHB. The maximum of the
spectra is also indicated (lines). This maximum has a
discontinuity at a point where the maximal value in the QP band
becomes larger than the maximal value reached in the LHB band.  Dashed
and solid lines are used to denote the maximum in the QP and LHB band,
respectively.  Above $T/D=0.2$ the maximal value does not have a
discontinuity anymore and the signature of the quasiparticles is only
visible as a kink in the dispersion. This marks the onset of the
bad-metal regime. Note that $T/D=0.2=\delta$ corresponds to the
Brinkman-Rice scale. At the highest temperature $T/D=1.0$ the kink is not
seen anymore.

The QP band crosses the Fermi energy at different momenta as
the temperature increases. Identifying the Fermi surface with the
momenta at which the spectral intensity of the QP band is maximal
leads to the conclusion that the Fermi volume inflates as the
temperature is increased. Note that the number of particles is fixed
so that with this identification of the Fermi surface the Luttinger
theorem is only obeyed at very low temperatures.

\begin{figure}
\includegraphics[width=1\columnwidth]{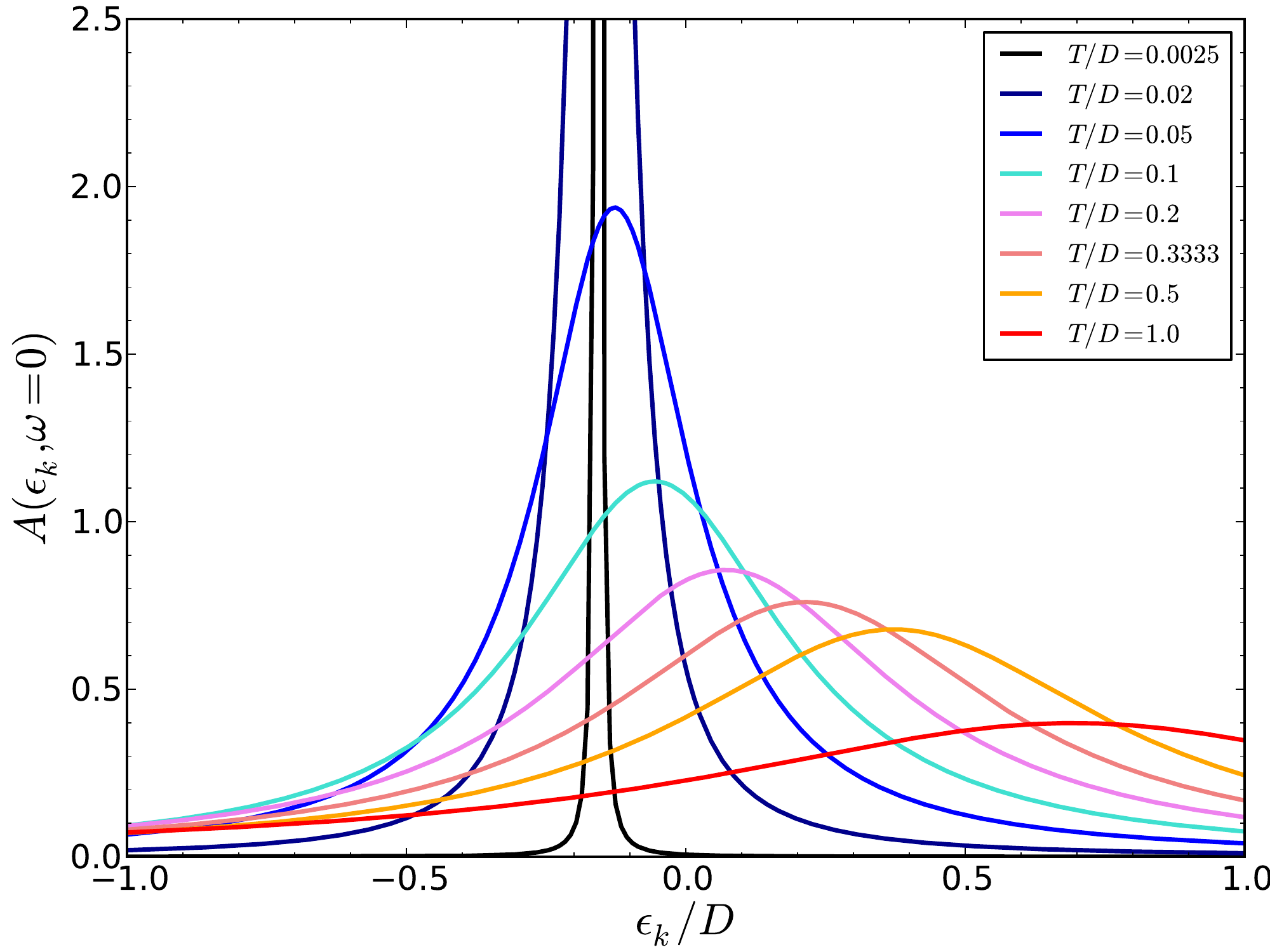}
  \caption{The evolution of momentum-distribution curves at the Fermi level Eq.~(\ref{eq:mdc}).
\label{fig:Ak0} }
\end{figure}

To elaborate on this, we plot on Fig.~\ref{fig:Ak0} the
momentum-distribution curve at $\omega=0$,
\begin{equation}
  A(\epsilon_k,0) =
    \frac{1}{\pi}\frac{-\mathrm{Im}\Sigma(0,T)}{(\mu-\mathrm{Re}\Sigma(0,T)-\epsilon_k)^2+(\mathrm{Im}\Sigma(0,T))^2} \label{eq:mdc}
\end{equation}
for several temperatures. This spectral function has the shape of a
Lorentzian centered at $\mu-\mathrm{Re}\Sigma(0,T)-\epsilon_k$ with a width
$\mathrm{Im}\Sigma(0,T)$.
At very low temperatures a sharp peak lies at the
chemical potential, fulfilling the Luttinger theorem.  When the
temperature increases, the peak moves to higher momenta. 

\begin{figure}
\includegraphics[width=1\columnwidth]{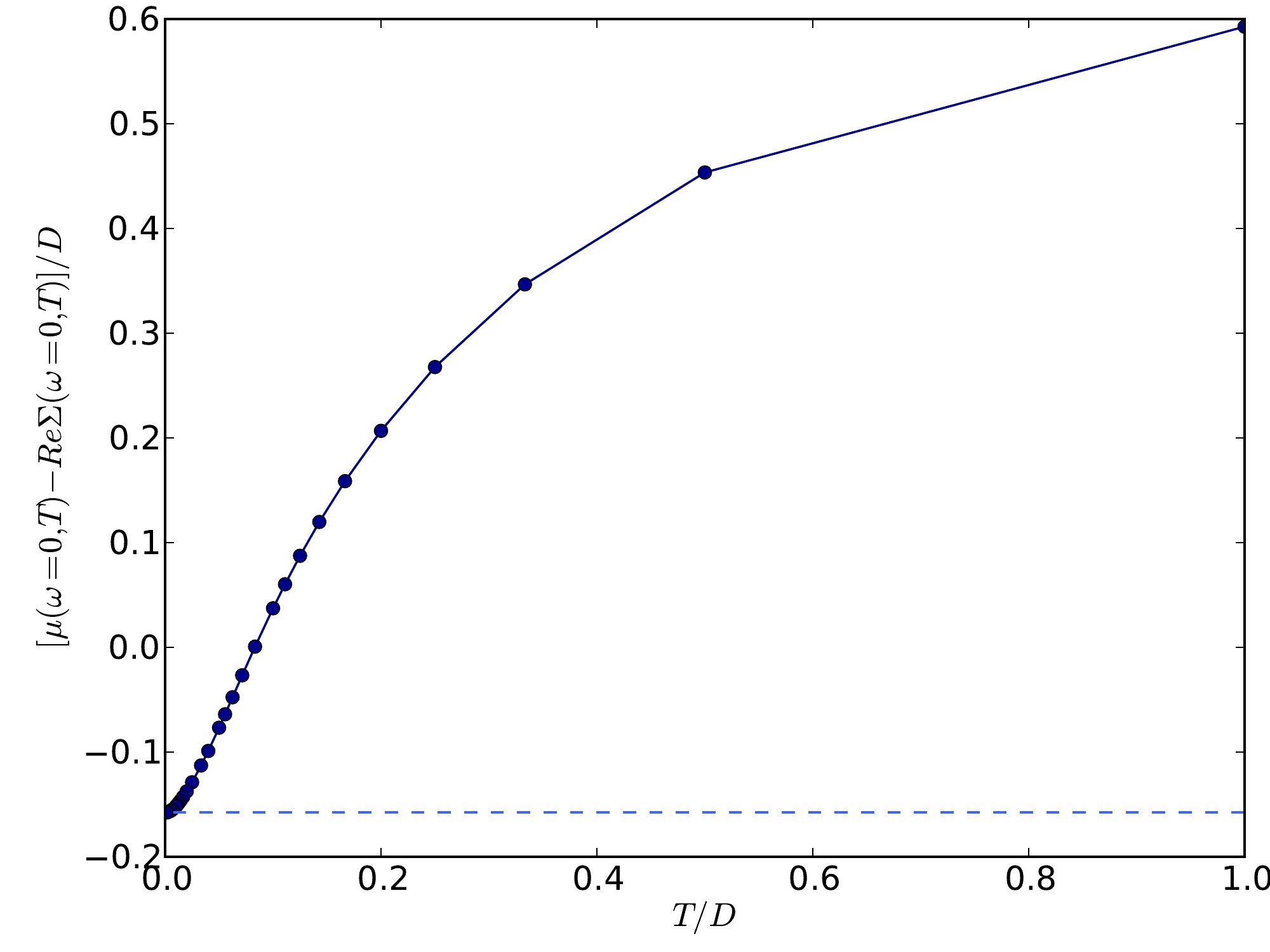}
  \caption{Temperature dependence of
    $\mu_\mathrm{eff}=\mu(T)-\mathrm{Re}\Sigma(\omega=0,T)$. The noninteracting
    chemical potential $\mu_0$ is shown by a dashed
    line.  \label{fig:mu_eff}}
\end{figure}

An
alternative way to track this change is to look at the renormalized
chemical potential $\mu_\mathrm{eff} = \mu-\mathrm{Re}\Sigma(0,T)$ as
a function of the temperature as shown in Fig.~\ref{fig:mu_eff}. In the
Fermi-liquid regime, $\mu_\mathrm{eff}$ essentially follows the
noninteracting chemical potential $\mu_0$ (shown with a dashed line). At
higher temperatures $\mu_\mathrm{eff}$ rapidly increases.

 An
important lesson here is that observing a well-defined QP
peak does not imply that the system has reached the Fermi liquid
regime. As Fig.~\ref{fig:mu_eff} shows, the chemical potential in this
intermediate-temperature metal can be quite far from the Fermi energy,
despite a signature of well-distinguishable resilient QPs.

\section{Thermopower at high temperatures and comparison to approximate formulas}

\begin{figure}
\includegraphics[width=1\columnwidth]{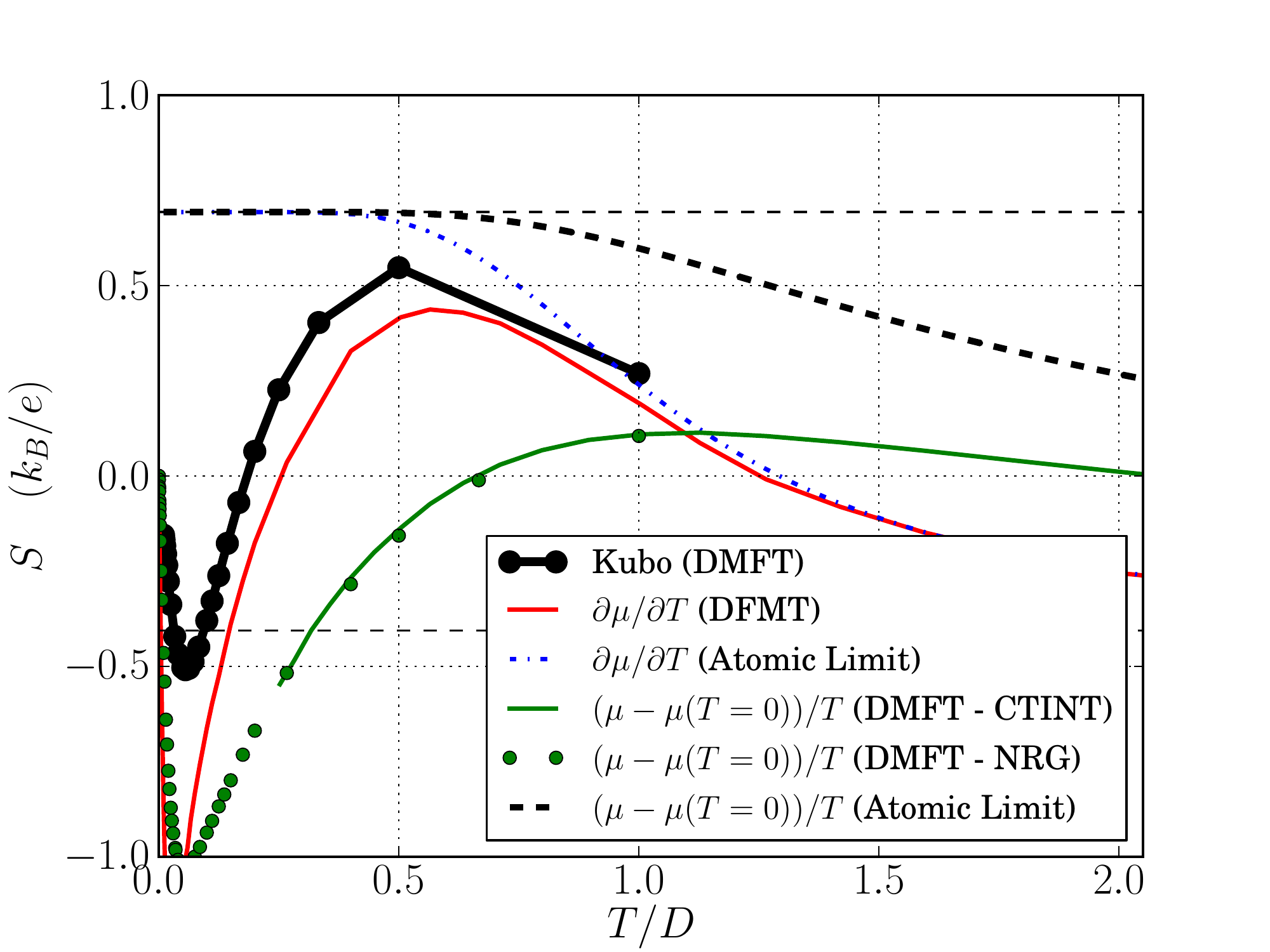}
  \caption{Seebeck coefficient calculated using the exact Kubo formula
    (thick black line) compared to the approximate Kelvin formula (red
    line) and Heikes formula (green line and green symbols). The
    atomic Kelvin estimate (dotted) and Heikes estimate (thick dashed)
    interpolate between the asymptotic $U\to\infty$ and $U\to 0$
    Heikes values.\label{fig:seebeck_highT}}
\end{figure}

In Fig.~\ref{fig:seebeck_highT} we show the Seebeck coefficient over a
larger temperature window. The Seebeck coefficient calculated using
the Kubo formula is plotted with a thick line and compared to various
estimates. The Kelvin formula $\partial \mu/\partial T$ (using $\mu$
calculated within DMFT) overestimates the magnitude of the thermopower in
the low-$T$ regime but is a good approximation of the exact result above
$T_*\approx 0.08D$. For comparison, we also plot the corresponding
atomic estimate using the Kelvin formula, but using the $\mu$
obtained in the atomic limit. We note that the two expressions
essentially match above $T/D=1$. Note that the Kubo result starts to
deviate significantly from the atomic estimate only when entering the
resilient QP regime.

Finally, we plot Heikes estimates. The results obtained from the
DMFT chemical potential using NRG and continuous-time interaction
expansion Monte Carlo (CTINT) are plotted (full green line and
symbols).  Heikes formula is found to approximate the thermopower
worse than the Kelvin formula. For comparison, also the atomic Heikes
estimates (thick dashed) as well as the asymptotic $U\to0$ and $U\to
\infty$ Heikes values (horizontal lines) are
shown.

\end{document}